\documentclass[acmsmall,screen,nonacm]{acmart}
\graphicspath{{figures/}}
\ExplSyntaxOn
\cs_new_protected:Npn \ps_todo_text:n #1
  {
    \group_begin:
    \tl_set:Nn \l_tmpa_tl {#1}
    \regex_replace_all:nnN { \b (G[0-9]+) \b }
      { \c{glref}\cB\{\1\cE\} } \l_tmpa_tl
    \tl_use:N \l_tmpa_tl
    \group_end:
  }
\ExplSyntaxOff
\newcounter{mysec}[subsection]
\counterwithin*{mysec}{subsubsection}
\newcommand{\mysec}[1]{\refstepcounter{mysec}\textbf{\themysec) #1}\enspace}
\newcommand{\mypara}[1]{\par\noindent\textbf{#1}\enspace}
\usepackage{array}
\usepackage{placeins}
\usepackage{longtable}
\usepackage{xurl}
\usepackage{tikz}
\usepackage{standalone}
\definecolor{glnavy}{RGB}{0,0,128}
\definecolor{pyellow}{RGB}{184,134,11}
\newcommand{\pquote}[2]{{\color{pyellow}``#1''}\if\relax\detokenize{#2}\relax\else~\textcolor{pyellow}{(#2)}\fi}
\expandafter\def\csname glurl@G1\endcsname{https://bib-gen.com/blog/i-built-a-race-bib-generator-for-my-own-birthday-run}
\expandafter\def\csname glurl@G2\endcsname{https://memoryofsnow.com/notes/hibiscus/}
\expandafter\def\csname glurl@G3\endcsname{https://www.reddit.com/r/lovable/comments/1qr1ixa/built_an_allinone_utility_app_with_zero_coding/}
\expandafter\def\csname glurl@G4\endcsname{https://medium.com/design-bootcamp/a-designers-first-attempt-at-building-with-claude-code-571686dfa17f}
\expandafter\def\csname glurl@G5\endcsname{https://olickel.com/my-home-cooked-meals}
\expandafter\def\csname glurl@G6\endcsname{https://zackproser.com/blog/ctx-personal-context-engine}
\expandafter\def\csname glurl@G7\endcsname{https://www.sebasab.com/blog/building-dream-journal}
\expandafter\def\csname glurl@G8\endcsname{https://lee-it.co.uk/software/2026/05/13/bubbyway-a-household-app-for-two.html}
\expandafter\def\csname glurl@G9\endcsname{https://www.kennethlacroix.me/blog/seven-months-of-vibe-coding/}
\expandafter\def\csname glurl@G10\endcsname{https://pkarnal.com/blog/building-my-ai-agent}
\expandafter\def\csname glurl@G11\endcsname{https://v5.chriskrycho.com/journal/build-yourself-small-tools/}
\expandafter\def\csname glurl@G12\endcsname{https://alexsci.com/blog/personal-apps/}
\expandafter\def\csname glurl@G13\endcsname{https://medium.com/@mudunuruchandu/i-built-a-personal-finance-tracker-that-fits-in-a-single-file-e5c2615d0404}
\expandafter\def\csname glurl@G14\endcsname{https://zerotrickpony.com/articles/software-for-one/}
\expandafter\def\csname glurl@G15\endcsname{https://www.reddit.com/r/ChatGPTCoding/comments/187pbs5/just_completed_a_632_line_ui_to_text_script/}
\expandafter\def\csname glurl@G16\endcsname{https://medium.com/@michaelakolackova/how-i-vibe-coded-a-village-tool-lending-library-cb254c29680b}
\expandafter\def\csname glurl@G17\endcsname{https://www.kuril.in/blog/the-last-carpenter/}
\expandafter\def\csname glurl@G18\endcsname{https://cstanton.net/posts/i-built-a-crm-twice-with-ai}
\expandafter\def\csname glurl@G19\endcsname{https://dehora.net/journal/2026/3/agentic-engineering-building-without-writing}
\expandafter\def\csname glurl@G20\endcsname{https://www.retiredpdx.com/p/building-a-gratitude-app-i-wanted}
\expandafter\def\csname glurl@G21\endcsname{https://jamiehurst.co.uk/2026-04-29_journal-v1}
\expandafter\def\csname glurl@G22\endcsname{https://yu-wenhao.com/en/blog/ai-meeting-notes/}
\expandafter\def\csname glurl@G23\endcsname{https://thesephist.com/posts/tools/}
\expandafter\def\csname glurl@G24\endcsname{https://huonw.github.io/blog/2021/08/home-cooked-app/}
\expandafter\def\csname glurl@G25\endcsname{https://iafisher.com/blog/2022/02/khaganate}
\expandafter\def\csname glurl@G26\endcsname{https://jakemccrary.com/blog/2020/10/03/go-create-silly-small-programs/}
\expandafter\def\csname glurl@G27\endcsname{https://tatey.com/2022/01/04/how-namestork-evolved-from-an-idea-to-a-tiny-service-over-four-years/}
\expandafter\def\csname glurl@G28\endcsname{https://www.robinsloan.com/notes/home-cooked-app/}
\expandafter\def\csname glurl@G29\endcsname{https://www.jefftk.com/p/designing-low-upkeep-software}
\expandafter\def\csname glurl@G30\endcsname{https://ezhart.com/posts/i-wrote-another-to-do-app}
\expandafter\def\csname glurl@G31\endcsname{https://simon.podhajsky.net/blog/2026/clobsidian-in-detail/}
\expandafter\def\csname glurl@G32\endcsname{https://ricardolara.dev/blog/build-your-own-cli/}
\expandafter\def\csname glurl@G33\endcsname{https://www.beginningwithi.com/2026/03/03/retirement-hobby-built-my-first-app-in-under-a-week/}
\expandafter\def\csname glurl@G34\endcsname{https://avandamiri.com/2009/07/16/piggy-back-launch-trials-and-tribulations.html}
\expandafter\def\csname glurl@G35\endcsname{https://cutedesk.app/blog/how-i-built-cute-desk-app-in-32-days.html}
\expandafter\def\csname glurl@G36\endcsname{https://blog.ewancroft.uk/3mm2rj3ejg22x}
\expandafter\def\csname glurl@G37\endcsname{https://blog.jread.com/posts/odyssey-projects/}
\expandafter\def\csname glurl@G38\endcsname{https://blog.kulman.sk/why-i-still-build-ios-apps/}
\expandafter\def\csname glurl@G39\endcsname{https://aetos-devlogs.hashnode.dev/building-webnotes-the-whole-story-so-far}
\expandafter\def\csname glurl@G40\endcsname{https://www.werkema.com/programming/the-spacemonger-1-x-post-mortem/}
\expandafter\def\csname glurl@G41\endcsname{https://blog.karannavani.com/how-i-built-and-launched-find-a-delivery}
\expandafter\def\csname glurl@G42\endcsname{https://tech.stonecharioteer.com/posts/2026/merrilin-local-sync/}
\expandafter\def\csname glurl@G43\endcsname{https://yakko.dev/blog/teeny}
\expandafter\def\csname glurl@G44\endcsname{https://booplet.com/blog/personal-apps/}

\newcommand{\glref}[1]{\href{\csname glurl@#1\endcsname}{\textcolor{glnavy}{#1}}}
\newcommand{\glquote}[2]{{\color{glnavy}``\emph{#1}''}~(\glref{#2})}
\usepackage{listings}
\lstdefinestyle{prompt}{basicstyle=\ttfamily\footnotesize, columns=fullflexible,
  keepspaces=true, upquote=true, breaklines=true, breakatwhitespace=true,
  breakindent=1em, xleftmargin=\leftmargini, xrightmargin=\leftmargini,
  aboveskip=6pt, belowskip=6pt,
  literate={…}{{\textrm{\ldots}}}1 {’}{{\textquoteright}}1}
\usepackage[capitalise,noabbrev]{cleveref}
\crefname{mysec}{finding}{findings}

\setcopyright{none}
\acmDOI{}
\acmISBN{}
\acmYear{2027}
\title{A Mixed-Methods Study of Personal Software Creation}
\author{Shaokang Jiang}
\email{shj@uci.edu}
\affiliation{%
  \institution{University of California, Irvine}
  \city{Irvine}
  \state{California}
  \country{USA}
}

\author{Kira Yuha Cho}
\email{kirac4@uci.edu}
\affiliation{%
  \institution{University of California, Irvine}
  \city{Irvine}
  \state{California}
  \country{USA}
}

\author{Emerson Murphy-Hill}
\email{emerson.rex@microsoft.com}
\affiliation{%
  \institution{Microsoft}
  \city{Seattle}
  \state{Washington}
  \country{USA}
}

\author{Daye Nam}
\email{daye.nam@uci.edu}
\affiliation{%
  \institution{University of California, Irvine}
  \city{Irvine}
  \state{California}
  \country{USA}
}

\begin{document}
\begin{abstract}
People have long wanted software shaped around their own data, routines, and
preferences. We refer to such software as \emph{personal software}. 
For decades, programming expertise limited who could build it: people without that expertise could build software only within constrained end-user programming tools, such as spreadsheets.
AI coding agents seem to remove this bottleneck by letting people build full applications from natural-language
descriptions, and many people now build and share personal software publicly.
Yet personal software has rarely been studied at scale: we know little about
what people build and why, how they build it, and what opportunities and
challenges they encounter.
We investigate these questions with a mixed-methods study. A grey-literature
review examines 6,168 web and Reddit posts, and an observation
study follows eight people as they build software for their own needs with AI
coding agents and use it for up to a week.
We find that creators value personal software for how closely it fits a particular person or community. The software takes shape as creators react to what the agent builds: most functionality emerged during building, and creators' effort shifted from writing code to checking it.
\end{abstract}

\maketitle

\section{Introduction}
\label{sec:introduction}

Since the early introduction of personal computers in the 1970s, computers have been envisioned as tools that individuals could shape to support their own thinking, problem solving, and everyday activities. Kay and Goldberg envisioned the personal computer as a programmable medium that could serve an individual's information and creative needs~\cite{kay1977personal}, building on Engelbart's view of computing as a means of ``augmenting human intellect''~\cite{engelbart2023augmenting}.
The vision of using computers to support individuals' activities assumed that people could adapt the software they used to the particulars of their lives: the information they keep, the routines they follow, and the tasks they want to accomplish.

For decades, this vision was only partly realized.
Commercial software is mostly designed for many users at once, encoding general assumptions about how a task should be done~\cite{litt_malleable_2025}.
Therefore, existing products have been addressing much of general users' need, while leaving a missing capability or workflows for individuals.
Users can tailor such products only within the options their designers provided~\cite{maclean_user-tailorable_1990}, and they often do not, because customizing takes time away from the work the software is meant to support~\cite{mackay_triggers_1991}.

The alternative is \textit{personal software}: software that people build primarily for their own use, shaped around their data, routines, and preferences.
Others have described similar software as ``software for one''~\cite{meiklejohn_software_2026}, a ``home-cooked meal''~\cite{noauthor_app_nodate}, ``home-cooked software''~\cite{noauthor_home-cooked_2025}, or ``situated software''~\cite{shirky2004situated}. 
People who could program, or who knew how to use end-user programming tools (e.g., trigger-action rules), built software for themselves~\cite{nardi1993small,ur_practical_2014,smith_build_2015,ko2011state}.
Until recently, people often could not build the software they envisioned~\cite{myers_invited_2006,ko2004six}, because building personal software required programming expertise. 

Generative AI seem to remove this bottleneck; as NVIDIA CEO Jensen Huang put it, ``Everyone is a programmer now---you just have to say something to the computer''~\cite{prakash_huang_2023}. 
Indeed, through natural-language interaction, people can ask an agent to generate an application and revise it, even without the programming experience to implement it themselves~\cite{chou2026building,sarkar_vibe_2025}. 
Many people now build personal software this way and share it on blogs, forums, and subreddits. Our search alone found about 30,000 public accounts of such software (\cref{sec:grey-literature-method}).
An era of personal software may be beginning, in which building software for oneself is no longer limited to programmers.

Despite its long history and recent visibility, personal software has rarely been studied as a category of software in its own right.
Prior research has examined it through particular tools, such as spreadsheets~\cite{nardi1993small} and trigger-action rules~\cite{ur_practical_2014}; through particular populations, such as developers building homegrown tools at work~\cite{smith_build_2015}; or through particular ways of building, such as vibe coding~\cite{chou2026building,sarkar_vibe_2025}.
These studies focus on how people program or prompt rather than on the software people build for themselves and its role in their lives.
We therefore lack a basic characterization of personal software: what people build and why, how they build it, and what opportunities and challenges they encounter. 

We organize our investigation around three research questions:

\begin{itemize}
    \item RQ1: What characterizes personal software and the needs it serves?
    \item RQ2: How do people create and evolve personal software?
    \item RQ3: What opportunities and challenges do creators encounter when building and using personal software?
\end{itemize}

To answer these research questions, we conduct a mixed-methods study with two parts. 
First, we conduct a grey-literature review that analyzes 6,168 web and Reddit posts by creators.
With the growth of personal software, many creators now describe and share their projects publicly, allowing us to examine diverse personal software at scale, built by diverse people in diverse domains.
However, public accounts include only what creators choose to report, and thus may not provide sufficient detail.
Thus, we conduct an observation study of eight people as they build software for their own needs with AI coding agents, walk through usage scenarios on the result, and optionally use it for a week. 
The creator accounts provide breadth and access to long-term use, while observation shows how participants express intent, respond to generated software, and evaluate its behavior.

From this mixed-methods study, we find that personal software is defined by how closely it fits a particular person or community rather than by what it does: creators built it because general-purpose tools did not fit their workflows, routines, or budgets, and its scope could stay bounded because meeting the need was enough (RQ1).
Personal software took shape through a loop of prompting, trying, and revising: most functionality (68\%) emerged during building rather than being planned, and creators' effort shifted from writing code to checking it, with 47\% of building time spent testing or reading the agent's output and 0.2\% editing code (RQ2).
Finally, with the entry barrier lowered by AI agents, creators learn through building personal software and gain additional value by sharing it with others.
At the same time, engineering work remains a challenge: simple-looking features hid hard problems such as data synchronization; upkeep and external dependencies required ongoing effort; and serving others required further work (RQ3).

Our contributions are:
\begin{itemize}
    \item A grey-literature review of 6,168 web and Reddit posts describing personal software, identifying the needs, values, and challenges creators report.
    \item An observation study of eight people building software for their own needs with AI coding agents, including a seven-day follow-up completed by six participants.
    \item A combined account of how situated needs become software, linking creator accounts with observed changes to functionality, verification practices, and experiences of continued use.
\end{itemize}

\section{Background and Related Work}
\label{sec:related-work}

\subsection{Personal Software}

Personal software has a long tradition in developer tooling: shell scripts, dotfiles, and personal utilities that programmers have always built for themselves~\cite{smith_build_2015, nardi1993small}. 
Building software for individual needs has a long history. Back in 1998, in the early days of open-source software, Raymond observed that ``every good work of software starts by scratching a developer's personal itch''~\cite{Raymond_1998}. This implies that some software is first built around its creator's own needs and conventions, and may later be shared with others who have similar needs~\cite{raymond1999cathedral,hippel2003open}.

Since then, personal software has emerged in multiple places. About 75\% of developers from Microsoft had built homegrown tools, which they develop on their own initiative to meet a personal need~\cite{smith_build_2015}; 84\% of scientists developed software tailored for their own research~\cite{hannay2009scientists}; Students are building ``situated software'' to rate professors for their own program's community~\cite{shirky2004situated}; and Sloan built ``a home-cooked'' photo and video messaging app for his family of four~\cite{noauthor_app_nodate}.
Nowadays, generative AI lets many more people, including non-programmers, build new and more personal software~\cite{roose_not_2025,karpathy_menugen_2025,meiklejohn_software_2026}, and many expect such software to proliferate~\cite{noauthor_home-cooked_2025,appel_personal_2025,pesce_vibe_2025}.
With this growing trend, the origins, motivations, evolution, and future integration of such software, as well as the software itself and the difficulties that arise in between, have not been studied, especially given the recent advances in AI development. Our study addresses this gap.

\subsection{End-User Programming and End-User Software Engineering}
Closely related field to personal software is end-user programming, which has been studied for decades.
That literature has examined non-professionals creating computational artifacts across spreadsheets~\cite{nardi1993small}, web mashups~\cite{wong_making_2007}, trigger-action rules~\cite{ur_practical_2014}, and scripting~\cite{nardi1993small}, establishing that they can and do create useful programs but that their artifacts are often error-prone and difficult to test, debug, and maintain~\cite{panko_what_1998,ko2011state,burnett2014future}. End-user software engineering (EUSE) responded by developing methods to help these creators produce more reliable artifacts: spreadsheet testing~\cite{liu_teaching_2010}, the What You See Is What You Test (WYSIWYT) methodology~\cite{rothermel_wysiwyt_2000}, assertion-based approaches~\cite{burnett_end-user_2003}, unit inference~\cite{erwig_adding_2002}, and formula auditing tools~\cite{burnett2004end}.
EUSE supports reliability by making authored complexity visible, assuming that creators can reason about the representation they manipulate~\cite{burnett2009end}:
spreadsheet users can inspect cells and formulas, and trigger-action programmers can read their rules. 
We inherit its population, anyone who wants to create their own program, and lifecycle concerns (e.g., how non-professionals test, verify, and maintain software built for their own needs).
However, as AI coding agents can generate full applications whose underlying code is difficult for non-experts to inspect~\cite{liu2023what,sarkar_vibe_2025}, we cannot assume that users can reason about the software,
which limits EUSE's inspection-based path to reliability.

\subsection{AI-Assisted Code Generation and Vibe Coding}

Early research on AI-assisted coding focused mainly on professional software development. Research on code-completion tools such as GitHub Copilot has examined developer productivity~\cite{paradis_how_2025, peng_impact_2023}, code quality~\cite{he_speed_2026}, and interaction strategies~\cite{barke_grounded_2023}. Studies have also identified difficulties understanding and debugging generated code~\cite{vaithilingam2022expectation}, challenges controlling assistants to obtain desired outputs~\cite{liang2024large}, and the substantial time spent evaluating suggestions~\cite{mozannar2024reading}. Even experienced developers do not always verify generated code~\cite{perry_users_2022}, despite evidence that suggestions can contain security vulnerabilities~\cite{pearce2025asleep} and that developers have difficulty appropriately validating and calibrating their trust in them~\cite{wang2024investigating}.

The landscape has since expanded beyond inline code completion to include agentic tools that can operate across repositories and generate larger portions of applications from natural-language descriptions~\cite{jimenez2024swe,yang2024swe}. 
Platforms such as Cursor\footnote{\url{https://cursor.com/}}, Claude Code\footnote{\url{https://claude.com/product/claude-code}}, Replit Agent\footnote{\url{https://replit.com/products/agent}}, Google AI Studio\footnote{\url{https://aistudio.google.com/}}, and Bolt\footnote{\url{https://bolt.new/}} allow users to describe what they want and have the system generate or modify substantial parts of an application, often without writing or reading any code. 
Karpathy coined the term \textit{vibe coding} to describe this practice: ``you fully give in to the vibes, embrace exponentials, and forget that the code even exists''~\cite{karpathy2025vibe}.

Empirical research on vibe coding is nascent. Chou et al.~\cite{chou2026building} studied how vibe coders manage the stochasticity of AI-generated code, identifying strategies such as undo, version control, small incremental changes, tests, and linters as safeguards against unwanted modifications. Other work has examined the prompting strategies~\cite{fawzy_prompting_2026, nam_understanding_2025} and failure modes~\cite{fawzy_prompting_2026, lin_is_2026} that arise when non-experts interact with coding agents. Studies of novice programmers found that generative AI can improve task completion~\cite{kazemitabaar2023studying} but can also compound existing metacognitive difficulties~\cite{prather2024widening}.

Our work differs from this literature in two respects. 
First, vibe coding research studies how people interact with coding agents during vibe coding, while we study what gets built, why people build such software, whether it works, and what happens next. 
Second, prior work has mostly examined developers in professional settings, practitioner accounts, or mixed-experience samples building general-purpose software~\cite{nam_understanding_2025,chou2026building,fawzy_prompting_2026}, rather than following people with varied programming experience as they build software for their own self-identified needs and then use it. 

\section{Study Overview}
\label{sec:study-overview}
    \begin{figure}
    \centering
    \includegraphics[width=0.8\columnwidth]{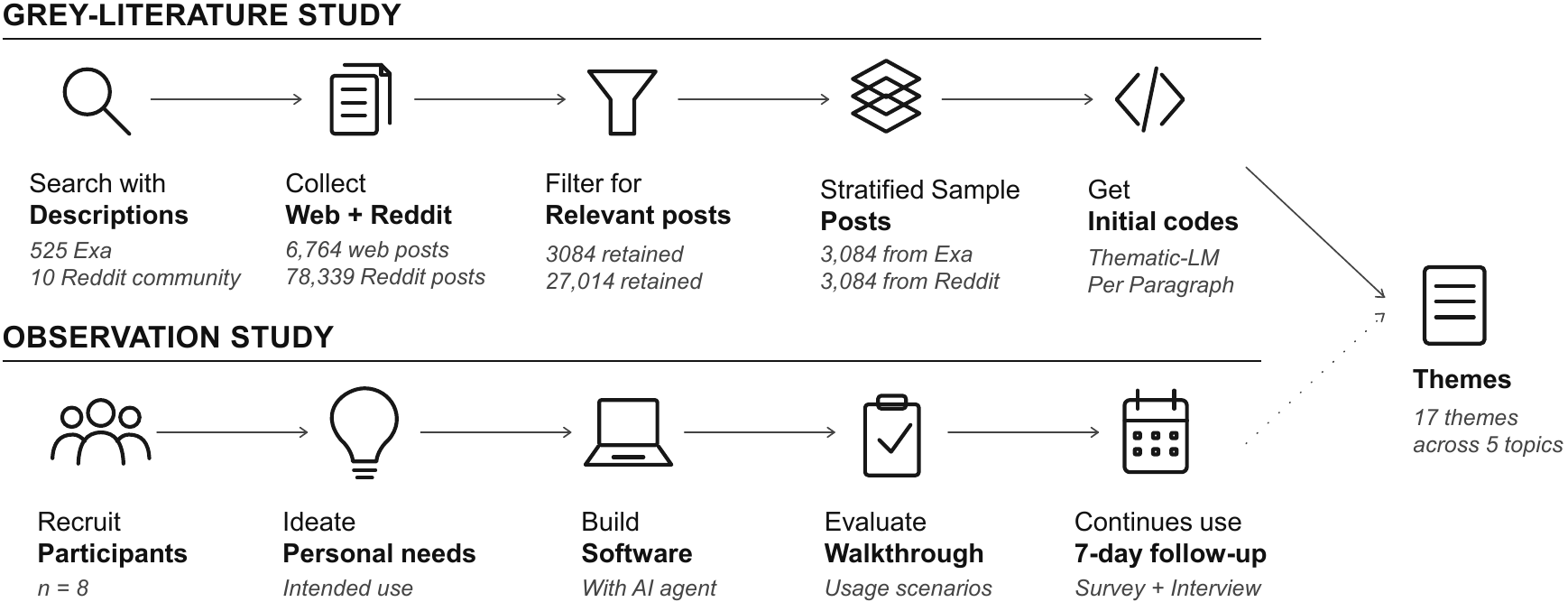}
    \caption{Overview of the two studies. The grey-literature review provides broad, retrospective accounts of personal software, while the observation study follows eight participants building and using AI-assisted personal software for up to a week.}
    \label{fig:study-overview}
\end{figure}

\begin{figure*}[t]
    \centering
    \includegraphics[width=\textwidth]{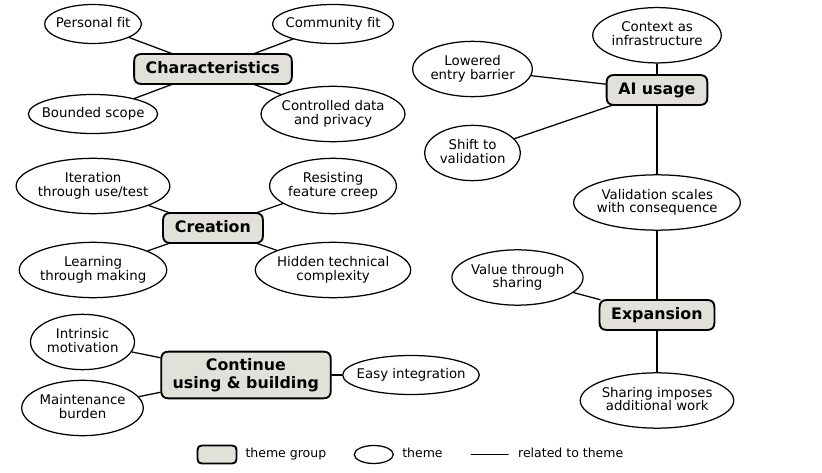}
    \caption{Theme groups (rounded rectangles) and themes (ovals) developed from
    the grey-literature corpus. Themes linked to two groups belong to both.}
    \label{fig:theme}
    \vspace{-\baselineskip}
\end{figure*}

We address our research questions with two complementary studies, as shown in \cref{fig:study-overview}. A grey-literature review provides broad, retrospective accounts of personal software across domains, populations, and stages of use.
An observation study follows eight participants as they build software for their own needs with AI coding agents and, optionally, use it for a week, capturing the creation and maintenance processes that public accounts may miss. Together, the studies provide breadth on personal software’s trajectory and direct evidence of how it is built and used.

We follow a convergent parallel mixed-methods design~\cite{creswell2017designing,storey2025guiding}: 
we conducted the two studies in parallel, analyzed them independently, and integrated the results at interpretation. 
Following a joint-display approach~\cite{fetters2013achieving}, we used the grey-literature themes (\cref{fig:theme}) as an organizing frame and compared the observation results against them. 

\subsection{Grey-literature review method}
\label{sec:grey-literature-method}
We screened public posts for accounts of personal software and conducted a grey literature review of 6,168 posts (3,084 web posts and 3,084 Reddit posts).
Creators described building tools to fit the needs of a particular person or community and refining them through use.
They reported that AI coding tools made it easier to start building, while testing, maintenance, and sharing still required substantial work.

We follow \citet{garousi2019guidelines} for the general grey-literature search, selection, and documentation guidelines, and adapted ~\citet{agarwal2026opinions} for running a scaled review workflow.
We combined semantic web search with Reddit posts, used LLMs for relevance filtering and initial coding, and refined LLM-generated candidate themes through team discussion.

\mysec{Data collection.} 
We collected web and Reddit posts about personal software.
For websites and blogs, we used Exa's semantic search\footnote{\url{https://exa.ai/}}.
Because people describe personal software in many ways (e.g., ``home-cooked software''), searching for the keyword ``personal software'' alone would miss most relevant posts.
We therefore first collected conceptual posts that introduce personal software and its variants, continuing until no new terms emerged, and extracted 75 descriptions and definitions from them (Appendix~\ref{sec:appendix-web-search} lists all descriptions).
We then paired each description with each of seven search prefixes (e.g., ``why I built'' and ``experience creating'') that the research team selected to target our research questions, yielding 525 queries for Exa.
With Exa semantic search of 50 results per query, we collected 6,975 unique web posts.
For paywalled or inaccessible pages, we recovered text from the original sites, archived versions, or Exa's cached content.
We excluded media pages without transcripts, leaving 6,764 web posts.

To complement web posts with first-person posts shared in discussion forums~\cite{proferes2021studying,reagle2022disguising}, we retrieved all available posts from the selected subreddits through the Arctic Shift archive~\cite{heitmann_arctic_shift}.
Based on team discussion, we selected 10 subreddits where people discuss or showcase software they built, covering both AI-assisted development and app sharing subreddits (AppIdeas, ShowYourApp, ShowMeYourApps, vibecoding, ClaudeAI, ChatGPTCoding, GitHubCopilot, lovable, GoogleAIStudio, and BuildWithClaude).
In five of these subreddits that provide clear tags for each post, we retained only posts with flairs that identified projects or showcases (details are in Appendix~\ref{sec:appendix-reddit-filtering}).
Across all 10 subreddits, we excluded posts with empty, deleted, or removed text, leaving 78,339 posts.

\mysec{Relevance filtering.}
Semantic search and collection from the selected subreddits also return off-topic posts, including conceptual essays, app ideas without an implementation, and descriptions of products built for a general market.
Because such off-topic posts could reduce the quality of the thematic analysis, we filtered the corpus for relevance.
Following prior LLM-assisted filtering and annotation approaches~\cite{agarwal2026opinions,ahmed2025can}, we developed a relevance classifier (\texttt{qwen3.6-35b-a3b-q6}) that classified each post three times and took the majority vote.
To build and evaluate it, human raters labeled 404 posts (297 web, 107 Reddit); two raters independently labeled 102 web posts ($\kappa=0.74$) before one rater labeled the rest.
We used 200 labeled web posts to refine the prompt and evaluated the classifier on held-out sets of 97 web and 107 Reddit posts, where it reached $\kappa=0.73$ and $\kappa=0.77$, respectively (Appendix~\ref{sec:appendix-relevance} in the supplemental material).
Filtering retained 3,084 web posts and 27,014 Reddit posts, totaling 30,098 relevant posts.
For analysis, we included all 3,084 web posts and sampled 3,084 Reddit posts, stratifying by subreddit to balance the two sources and keep coding cost manageable.

\mysec{Thematic analysis.} 
We conducted an inductive thematic analysis~\cite{braun2006using}
to characterize the range of needs, practices, and challenges creators of personal software describe. 
Given the size of our corpus and prior work using LLMs to support inductive qualitative coding and thematic analysis~\cite{gao2024collabcoder,dai2023llm,de2024performing}, we generated initial codes using Thematic-LM~\cite{qiao_thematic-lm_2025} with \texttt{gpt-6-luna}.
We coded at the paragraph level in batches of 100 units, 259,931 units in total.
Each unit was analyzed from three prompted perspectives: a software engineer who builds personal tools; an HCI and software engineering researcher; and a nontechnical maker. 
Given our large codebook and Thematic-LM's difficulty generating cohesive themes at this scale, we adapted the prompt of Agarwal et al.~\cite{agarwal2026opinions}, which asks the LLM to make each theme synthesize patterns across multiple codes and constrains the number of candidate themes. We set the cap at 50 candidate themes (Appendix~\ref{sec:appendix-coding-pipeline} in the supplemental material). The research team determined the final structure through 35 hours of discussion, manually refining, merging, and splitting candidate themes. The result contains 17 themes in five theme groups (\cref{fig:theme}).

\subsection{Observation Method}
\label{sec:observation-method}

To understand how people turn personal needs into working software, we observed eight participants build with AI coding agents and test their tools in their own usage scenarios.
In addition, for six participants who opted in, we conducted a seven-day follow-up study to understand how they use, maintain, and evolve their personal software.
Most functionality emerged as participants tried and revised what the agents built.

Our institution's IRB approved the study. The full protocol and further procedure details are in
Appendix~\ref{sec:appendix-observation} in the supplemental material.

\begin{table}[htbp]
  \centering
  \footnotesize
  \setlength{\tabcolsep}{3pt}
  \caption{Participant information.}
\begin{tabular}{
    @{}l
    l
    >{\raggedright\arraybackslash}p{2.7cm}
    >{\raggedright\arraybackslash}p{8.1cm}
    l@{}
}
    \toprule
    ID & Age & Built Software & Software Summary & Tool \\
    \midrule

    P1 & 18--24 & Emotion Tracker
    & Mood tracker with daily check-ins and trend visualization.
    & AI studio \\

    P2 & 18--24 & Mental Health Journal
    & Mental-health journal with check-ins, resources, trends, and cloud sync.
    & AI studio \\

    P3 & 18--24 & RoutePilot
    & Multi-stop route planner with per-stop ETA and navigation.
    & Claude \\

    P4 & 25--34 & Household Coordinator
    & Household app for groceries, expenses, chores, and calendars.
    & Claude \\

    P5 & 18--24 & FlashFlow Studio
    & Flashcard app with spaced repetition, Markdown, and cloud sync.
    & AI studio \\

    P6 & 18--24 & Resume Tailoring
    & ATS resume tailoring with keyword matching and PDF export.
    & Antigravity \\

    P7 & 18--24 & Envdex
    & Terminal dashboard for managing project environment variables.
    & Claude \\

    P8 & 45--54 & V-CAM
    & Delirium decision-support tool with anonymized assessment logging.
    & AI studio \\

    \bottomrule
\end{tabular}%
\label{tab:participants}
\vspace{-\baselineskip}
\end{table}

\subsubsection{Participants}

We recruited participants through social media, online forums, email lists, personal networks, and posters around campus. 
To include people both with and without programming experience, we advertised our study to diverse population, including a Slack workspace for neuroscientists and a Reddit for local community.
A prescreening survey asked for a situation in which the respondent wanted
personal software, their programming and agentic-coding experience, and
demographics. We prioritized respondents with domain experience and a clear idea
of what to build, and did not recruit anyone whose idea duplicated an earlier
participant's. Compensation was \$15 per hour for the Session~1 interviews, \$10
per hour for building, \$5 for the seven-day survey, and \$20 per hour for the
follow-up interview. \cref{tab:participants} summarizes the participants and
their software. Three pilot sessions, excluded from the analysis, revealed that participants often declared their software complete before it was usable. We therefore added a scenario walkthrough, asking participants to test their software against their identified scenarios and identify any immediate issues that needed fixing. (\cref{sec:appendix-observation}).

\subsubsection{Procedure}\mbox{}\par

\mysec{Session 1: Personal software creation.} The session had four parts.
(a)~In \emph{idea exploration} (10--15 minutes), participants described the software
they wanted, its purpose, how often they needed it, their experience with
existing software, and how they expected to use it. Although there was no time limit, all participants finished within 15 minutes.
 (b)~In \emph{building} (1--3
hours), they built a minimal viable product with their preferred AI coding tool,
or with Google AI Studio if they had none. There was generally no time limit, and participants could pause at any time. 
When participants asked us for help, we directed them to the agent or to a general-purpose chat assistant. We intervened only when a participant was stuck for more than three consecutive prompts, and then only by suggesting a meta-prompt.
Participants shared their screens and thought aloud~\cite{boren2000thinking}.
While the agent was generating a response, we asked participants about the intent behind their prompt and what they expected the agent to produce.
(c)~Once participants declared that they had a minimal viable product, they began the \emph{scenario walkthrough}: they wrote at least three usage scenarios following a modified user-story template~\cite{lucassen2018jobs, amna2022ambiguity, cohn2004user} that included the device and context of use.
Participants then tested each scenario on the  software while completing an evaluation survey, and could fix core issues the walkthrough exposed after completing the survey. 
(d)~In the \emph{closing interview} (10--15 minutes), we asked about their experience, motivation, development process, and the origin and evolution of the functionality.

\mysec{Session 2: Seven days of use.} 
After Session 1, participants who opted in were asked to use their software at least three times over seven days and keep a record of each use and any prompts~\cite{bolger2003diary}. 
A survey at the end of the week asked about usage and value, and included usability items adapted from the PSSUQ~\cite{lewis1992psychometric} and study-specific items on reliability, personal fit, ownership, and intention to continue.

\mysec{Session 3: Follow-up interview.} 
Participants who changed their software during Session 2 were invited to a follow-up interview within five days.
We asked them to demonstrate the current, update version, and to describe how it fit their intended activities, and what they changed, why, and how.

\mysec{Data collection.} We collected Zoom recordings, built personal software, survey responses, and prompts used during Session 1.
All participants reviewed shared materials to exclude sensitive information.

\subsubsection{Data Analysis}
\label{sec:obs-analysis}
To analyze the data, we mainly conducted iterative qualitative coding~\cite{saldana2009coding}.
To answer three research questions, we analyzed participants' transcripts, prompts, and video recordings, focusing on different questions:
\begin{itemize}
    \item participants' transcripts focusing on their \emph{motivation, opportunity, and difficulty} (RQ1),
    \item prompts focusing on \emph{functionality evolution} (RQ2),
    \item video recordings focusing on \emph{on-screen activity} (RQ2), and
    \item participants' transcripts focusing on \emph{opportunity and difficulty} (RQ3)
\end{itemize}

\mysec{Motivation, opportunity, and difficulty.}
For motivation, opportunity, and difficulty, we conducted an inductive,
iterative thematic analysis of participants' transcripts and corresponding
video recordings~\cite{braun2006using,nowell2017thematic}. 
Two researchers independently open-coded transcripts from the first two participants at \emph{meaning units}~\cite{graneheim2004qualitative} level, defined here as contiguous excerpts about a single relevant topic.
Each unit contained at least one complete sentence and could span paragraph boundaries.
When necessary, the researchers consulted the recordings for context. 
The researchers compared annotations and memos, resolved disagreements through discussion, and refined the codebook. 
Both then coded the remaining six participants, iteratively reviewing emerging codes and
interpretations. 
For each code assigned to a participant, we documented a memo explaining the rationale and referencing supporting evidence across the session for later discussion. 
A final round of discussions reviewed newly identified codes and uncertain cases and established consensus on the final assignments for each participant.

In this paper, we report the number of participants with a final consensus assignment, counting each participant once per code.
Due to space limits, we highlight selected codes and subcodes here.
Appendix~\ref{sec:appendix-obs-codebooks} in the supplemental material lists and defines all motivation, opportunity, and difficulty codes and subcodes.

For motivation, six general codes capture why participants wanted their software: \emph{personal
life/routine}, \emph{existing software gap/friction}, \emph{better supporting
others}, \emph{learning AI-assisted development}, \emph{resource constraints},
and \emph{translating domain expertise into software}. 
Three additional codes distinguish forms of existing-tool friction.
These are \emph{using multiple software tools for one task},
\emph{difficulty communicating and obtaining changes to an existing tool}, and
\emph{limited awareness or unwillingness to find existing tools}.

Regarding opportunities and difficulties while making personal software, codes fall into three categories, 
\emph{creator--platform interaction challenges}, \emph{difficulties using and
managing the generated software}, and \emph{opportunities afforded by the
AI-assisted development environment}.

\mysec{Functionality evolution.} 
To understand how personal software evolves during building (RQ2), we traced which capabilities participants planned in advance and how their requests changed as they built.
We devine a \emph{functionality} as a distinct software capability, such as archiving deleted journal check-ins (P2) or syncing flashcards across devices (P5).
Because one prompt could contain several requests, we split prompts into
sub-requests and grouped them by functionality.
An LLM (\texttt{qwen3.6-35b-a3b-q6}) split the prompts and rated the similarity of every pair of sub-requests, and we clustered the sub-requests with agglomerative hierarchical clustering~\cite{murtagh2012algorithms}, choosing the linkage method and number of clusters per participant by the highest silhouette score
(Appendix~\ref{sec:appendix-grouping} in the supplemental material). This yielded 131
functionalities (6--27 per participant) covering 396 sub-requests.

We ordered each functionality's sub-requests by time and labeled each with one of nine request types, by how it changed the functionality from earlier sub-requests. Seven are
\emph{build requests}, which add, change, or drop functionality
(\emph{Introduced}, \emph{Expanded}, \emph{Refined}, \emph{Narrowed},
\emph{Repaired}, \emph{Restated}, and \emph{Abandoned}).
The other two are \emph{inquiries} (\emph{Explanation} and \emph{Confirmation}).
We call each pair of consecutive labels within a functionality a \emph{transition} (e.g., \emph{Introduced} followed by \emph{Expanded}), representing one step in the functionality's evolution.

Beyond labeling each request, we also coded each functionality on whether it was planned in advance based on participant's initial plan in the Session~1 transcript.

Two researchers independently coded functionality planning and evolution, reaching substantial agreement ($\kappa=0.80$).
We refined the LLM coding procedure on one participant whose labels the researchers had agreed on, then evaluated it on a held-out participant and reached substantial agreement between human and LLM ($\kappa=0.701$).
We used LLM annotations for the remaining six participants.

\mysec{On-screen activity.} 
On-screen activity captures what participants did during building, including periods of waiting for the agent.
We analyzed it to understand how participants spent their time and moved between requesting changes and checking the results (RQ2).
We labeled two-second intervals of the building-phase videos with activities such as \emph{testing / reading}, \emph{waiting}, \emph{prompting}, \emph{configuring}, and \emph{editing code}, adapted from \citet{mozannar2024reading}.

To understand what participants verified, we coded which functionality was being verified in each interval labeled \emph{testing / reading}.
We counted a functionality as verified if, in at least one such interval, it appeared on screen and the participant was actively checking it.

We manually labeled the building-phase videos of P1 and P3 and used Cohen's $\kappa$ to compare the human and LLM activity labels on the same two-second intervals (pooled $\kappa=0.753$).
To validate the functionality being verified during \emph{testing / reading} session, we compared, for each participant, the set of functionalities the human marked as verified with the set the LLM marked as verified. We measured their overlap with Jaccard similarity~\cite{maier2024metrics,vanbelle2025measuring}: the number of functionalities in both sets divided by the number in either set. The mean Jaccard similarity between human and LLM over P1 and P3 was 0.83.
We used LLM annotations for the remaining six participants (Appendix~\ref{sec:appendix-llm-coding} in the supplemental material).

\section{Findings}
\label{sec:findings}

We report our findings under each research question, drawing on both studies.
\cref{fig:theme} shows the theme groups and themes developed from the grey-literature corpus. 
Each theme heading lists its evidence: \emph{grey-lit} reports the count and percentage of posts in the grey-literature review supporting that theme, out of 6,075 coded posts (3,018 Reddit and 3,057 web)\footnote{The other 66 sampled Reddit posts and 27 web posts had no validated code assigned to a final theme.}; \emph{obs} marks themes that the observation study also informs. Each post counts once per theme or RQ; \emph{Other} is excluded and percentages can overlap.
Each research question draws on themes from several groups, and each finding below corresponds to one theme. 
The titles and original links of the quoted grey-literature sources are listed by source ID in Appendix~\ref{sec:appendix-grey-literature-sources} in the supplemental material.
Quotes from the grey-literature review appear in {\color{glnavy}\emph{navy italics}}, followed by the source ID (e.g., \glref{G28}) linking to the original source; quotes from the observation study appear in {\color{pyellow}dark gold}, followed by the participant ID (e.g., \textcolor{pyellow}{P3}).

\subsection{RQ1: Personal Software Fits Individual Needs Within a Limited Scope}
\label{sec:rq1}

RQ1 asks what characterizes personal software and the needs it serves. Understanding these characteristics helps explain why people build their own software, why it is valuable to a particular person or community rather than a broad market, and why existing solutions may not suffice. Our six RQ1 themes cover 5,469 posts (90.02\%), including 2,526 unique Reddit posts and 2,943 unique web posts. Additionally, we supplement these accounts with interviews with eight observation-study participants about their needs, limitations of existing tools, and intended contexts of use.

In both studies, personal software was defined by how closely it fit a
particular person or community rather than by what it did. Creators built it
because general-purpose tools did not fit their workflows, routines, budgets,
or the people around them, and they designed it to work with the tools and
devices they already used and to keep control over their data. Its scope could
stay bounded: meeting the specific need could be enough. Beyond utility,
creators also built for enjoyment, learning, and creative expression.

\mysec{Personal fit (grey-lit: 3,534 posts, 58.17\%, obs).}
Personal software creates value by fitting specific lives and circumstances that general-purpose tools do not serve well.
Many creators built it to fit their own workflow,
\glquote{I want things to fit my workflow instead of adjusting my workflow to fit someone else's ideas}{G11},
including its temporal assumptions:
\glquote{At midnight Todoist would mark tasks for the current day as “overdue” and start showing the new day's tasks}{G12},
whereas that creator wanted the day to end when their own late-night routine ended. 
Cost could also be a personal requirement that existing products did not meet: \glquote{YNAB is excellent but costs \$109 a year}{G13}.

In the observation study, seven of eight participants named applications that existed but required inconvenient workflows, contained unnecessary complexity, or assumed a different way of doing the task, and four cited funding or budget constraints.
Map applications, for example, do not
\pquote{let you have an ETA or multiple stops when you're doing a walking route, even though you can do it for driving}{P3}.
The tasks themselves were rarely novel: most fell into established software categories, such as planning and tracking (\cref{tab:participants}). 
What made the software personal was how closely it fit the participant.

\mysec{Community fit (grey-lit: 1,653 posts, 27.21\%, obs).}
People also create software to
\glquote{make the lives of the people around me better}{G5}.
Some built software for their partner,
\glquote{makes a thing easier for my partner specifically}{G8}, 
or for a shared place or practice, such as a village tool lending library (\glref{G16}), whose value may depend more on local trust and relationships than on broad reach.
A wedding-countdown watch face showed shared photos and messages to the creator and their wife, deliberately serving a
\glquote{specific set of people: me and my wife}{G24}.

In the observation study, four participants wanted their software to support other people, such as roommates sharing a household (P4) or colleagues in a clinical team (P8).

\mysec{Controlled data and privacy (grey-lit: 2,705 posts, 44.53\%).}
Creator accounts treated control over data and behavior as part of what makes software personal, and losing that control could motivate building.
One creator found that \glquote{even exporting my data was treated as a premium capability}{G20} in a commercial app the creator had been using.
Many accounts describe local-first or privacy-first designs~\cite{kleppmann2019local}.
For example, the creator of Dream Journal stored entries in the browser's IndexedDB to leave
\glquote{no question about whether a private journal entry is being sent somewhere else}{G7},
and another creator ran a personal suite locally because \glquote{it's simpler and more secure than hosting it on the internet}{G25}.

\mysec{Intrinsic motivation (grey-lit: 1,452 posts, 23.90\%, obs).}
Intrinsic motivation could sustain personal software development beyond the
initial decision to build. Learning, enjoyment, creative expression, agency,
and making something useful for oneself or others could make continued work
worthwhile without requiring a large audience or a commercial goal, similar to 
the motivations for open-source contributors~\cite{lakhani_why_2003}.
For example, the creator of Upsilon, initially built to monitor his own servers
and messages, described extending and rewriting it over roughly a decade as
an opportunity for continued learning:
\glquote{I’ve invested THOUSANDS of hours in Upsilon, and I’ve learned SO much from it}{G37}.
The creator of a scrobble importer and a converter for recovering his own
content as Markdown described
\glquote{software that I maintain for no reason other than the fact that I still find it interesting}{G36}.
A creator of personal Japanese-reading, anime-tracking, and expense-recording
apps described fixing annoyances encountered in everyday use. Despite being
almost their only user, he concluded,
\glquote{At this point, that is reason enough to keep building}{G38}.

In the observation study, two participants named learning AI-assisted
development as a motivation; because we recruited people with a specific need,
the study was not designed to capture enjoyment-driven building.

\mysec{Easy integration (grey-lit: 2,549 posts, 41.96\%, obs).}
Creators built personal software to fit into the workflows, devices, and services they already used.
For example, a creator frustrated with Google Calendar built their own calendar so that it \glquote{appears on Khaganate's home page, which I check every morning}{G25}.
A household chore app ran on the kitchen iPad, so checking chores did not require another device (\glref{G12}).

In the observation study, participants used their software during the week when it fit into what they were already doing.
P1 filled in the emotion tracker's check-ins mostly when already on the laptop for homework rather than at bedtime, and because \pquote{I'm not usually, like, on my laptop}{P1}, expected to keep using it only sporadically.
By contrast, P3 kept the route planner for timing itineraries alongside Google Maps, because it could plan walking and driving legs \pquote{all in one go as opposed to Google Maps requiring me to do multiple inputs}{P3}, and planned to keep using it.

\mysec{Bounded scope (grey-lit: 683 posts, 11.24\%, obs).}
Personal software can stay small because it is bounded by the need it serves, and meeting that need can be enough for its creator to regard it as complete.
A minimal blog generator does little,
\glquote{All it does is generate pages based on HTML templates and Markdown content}{G43},
yet it gives its creator \glquote{a blog I'm happy with}{G43}.
Another creator saw an end point in such tools,
\glquote{If you scope them small enough, there is a definitive ending point}{G26},
and a writing-app creator whose app already met the need had
\glquote{no intention of adding more features to it}{G44}.

In the observation study, some participants were satisfied once the software covered the need it was built for. 
At the end of the session, P7 was 
\pquote{pretty satisfied with the current software}{P7}, which
\pquote{does exactly what it intends}{P7}, and after a week of use, P3 made no
changes because \pquote{the software's current capabilities were sufficient for my needs at the time}{P3}.

\subsection{RQ2: Personal Software Takes Shape Through Use and Iteration}
\label{sec:rq2}

RQ2 asks how people create and evolve personal software; it draws on five themes covering 4,922 posts (81.02\%; 2,199 Reddit and 2,723 web posts).
In both studies, 
most functionality was not planned in
advance: in the observation study, 89 of the 131 functionalities (68\%) emerged
during building. When AI agents wrote the code, checking the agent's work and planning the next iteration took 47\% of creators' time, while writing and fixing code by hand almost disappeared (0.2\% of time). How much
creators checked depended on what an error could affect, and they kept their
software focused by resisting features that did not serve their needs. Across
sessions, creators also had to preserve the context that made the software
personal.

\mysec{Iteration through use/test (grey-lit: 3,230 posts, 53.17\%, obs).}
Personal software starts from a specific need (\cref{sec:rq1}), but many of its requirements were discovered or refined only once creators used or tested it.
Creator accounts described needs emerging from use,
\glquote{As I used it, real needs emerged}{G32}, and from testing with real data,
\glquote{One of many features that I realized the need for as I started to test with real data}{G33}.

\begin{figure*}[t]
\centering
\includestandalone[width=\textwidth]{graph_gen/timeline_activity}
\caption{Every recorded session. Each row is one participant,
labeled with the software they built. Shading marks the session phases; during building, each two-second interval is colored by annotated on-screen activity. The evaluate phase covers the scenario walkthrough and evaluation
survey; in the refine phase, participants could fix issues the walkthrough
exposed.}
\label{fig:timeline}
\vspace{-\baselineskip}
\end{figure*}

\begin{figure}[t]
    \centering
    \includegraphics[width=\columnwidth]{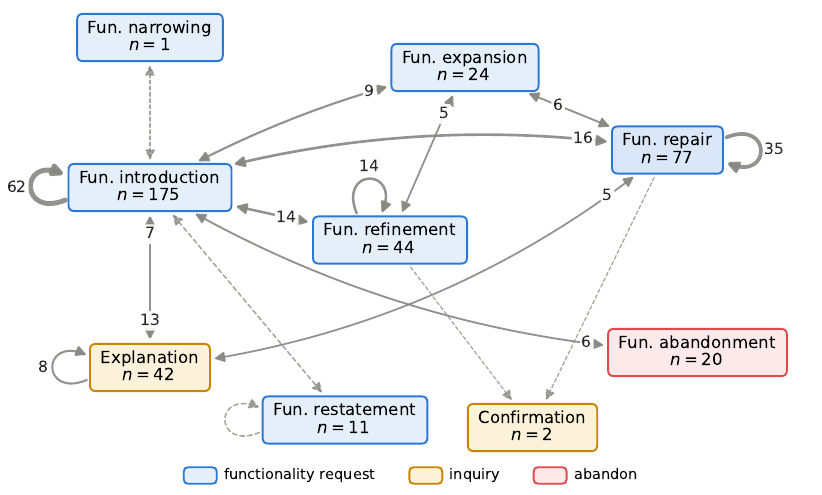}
    \caption{Transitions between request types during building. Nodes are the
    types of requests participants made about a functionality (Fun.), and $n$ is
    the number of requests of each type. \emph{Build requests} add, change, or drop functionality, and \emph{inquiries} ask the
    agent to explain or confirm. Edges are the main transitions from one request to the next about the same functionality, with counts shown for the frequent ones. Dashed edges are less frequent than solid ones. Double-headed arrows show both directions.}
    \label{fig:functionality_evolution}
    \vspace{-\baselineskip}
\end{figure}

\FloatBarrier
In the observation study, participants tried the software on their own situations while building it (\cref{fig:timeline}). 
Of the 131 functionalities participants worked on, 42 were planned before building and 89 (68\%) emerged during it.
P7, for example, added a way to jump to a project's directory after testing the tool:
\pquote{I experienced my own pain points, so I was having thoughts, like, maybe I want to open this project in its own directory}{P7}.
The scenario walkthrough after building revealed more: participants recorded something to change for 20 of the 28 scenarios, as when P1 found no way to \pquote{go back in time to check past dates}{P1} and added date navigation.

Functionality also rarely took shape in a single request (\cref{fig:functionality_evolution}): 103 of the 131 functionalities (79\%) involved multiple requests, with a median of 2 requests per functionality (range: 1--15).
Following the sequence of requests about each functionality (\cref{sec:obs-analysis}), we found that participants specified functionality piecemeal: 
the most frequent transition was from one introducing request to another for the same functionality (62 of 265 transitions), as participants added more detail across several requests. Repairs also came in runs: 37 of the 131 functionalities received at least one repair request, and a repair was most often followed by another repair (35 transitions), often restating the same failed behavior, so fixing a problem frequently took several attempts.

\mysec{Shift to validation (grey-lit: 2,196 posts, 36.15\%, obs).}
Building personal software with AI shifts the creator's effort from implementation toward validation: specifying goals, reviewing changes, and testing behavior. 
A working result did not end this work. One creator wrote that
\glquote{I could have stopped at 4 hours and had a working program, but I wanted more options, so I spent the next 8 hours tweaking and testing}{G15},
and another tested a tool lending library over multiple rounds of simulated usage scenarios by \glquote{clicking everything to see if it behaved as expected}{G16}.

In the observation study, participants spent most of the building phase checking the software: 
47\% of the annotated intervals showed participants \emph{testing} the software or \emph{reading} the agent's output, 28\% \emph{waiting}, 20\% \emph{writing prompts}, and 0.2\% \emph{editing code} by hand (\cref{fig:timeline}). During the session, participants tested and validated 93 of the 131 functionalities (71\%) they had explicitly asked for in their prompts.
86\% of \emph{waiting} periods was followed by \emph{testing} or \emph{reading}. 
This is consistent with prior work on AI-assisted programming, which found that programmers spend much of their time verifying suggestions~\cite{mozannar2024reading}.
However, finding a problem did not mean knowing how to fix it. 
All eight participants encountered a mismatch between expected and delivered behavior, and six repeatedly prompted to obtain a requested change:
\pquote{right now, everything is broken, and I'm like, I don't even know what to tell it or give it}{P5}.
Creator accounts similarly described repair introducing further mismatches,
\glquote{Each fix regressed something else. The agent thought it was done, but my eyes told me it wasn't}{G17}.
Relying on validation alone also had limits. Three participants had difficulty telling what the AI had done or changed, and two initially assumed a task was complete but later found that it was not.
Even with those validations, when participants walked through the usage scenarios they had designed right after declaring a minimal viable product, they judged only 12 of the 28 scenarios completely supported by the software they had just built.

\mysec{Validation scales with consequence (grey-lit: 1,680 posts, 27.65\%, obs).}
Creators weigh what an error would cost before relying on their software: the more its outputs affect their health, money, data, or other people, the more cautious they are and the more they verify first.
Creator accounts tied precautions to what an error could affect. 
The creator of a debt-sharing tool for their lunch companions prioritized the balancing algorithm because 
\glquote{no one would use it if it didn't crunch the numbers correctly, including me}{G34},
and worked through possible debt situations on paper before relying on it.
A personal-finance AI assistant required that a
\glquote{Human reviews reasoning and decides YES/NO}{G10} before any trade, and
a personal-information assistant had
\glquote{read-only access to local copies of my data --- no credentials that could delete my email or post on my behalf}{G31}.

\begin{figure*}[t]
\centering
\includegraphics[width=.85\textwidth]{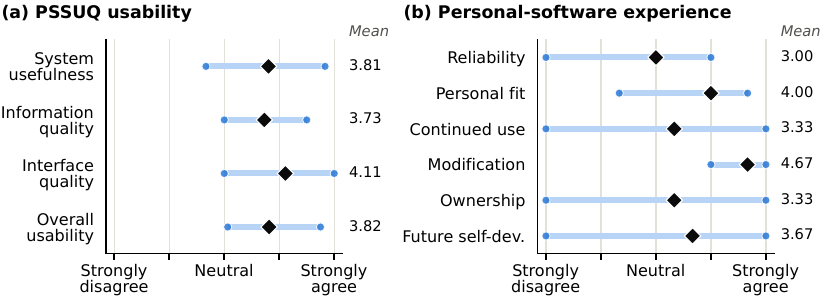}
\caption{Participants' perceptions of their software after seven days of use
($n=6$). Panel (a) shows PSSUQ-derived usability
measures; panel (b) shows study-specific measures
(Future self-dev.: wanting to build more software like this for oneself).
Means are shown at right (1 = strongly disagree, 5 = strongly agree).}
\label{fig:perception}
\vspace{-1\baselineskip}
\end{figure*}

In the observation study, participants were cautious where an error could cause harm. 
P8, whose tool estimated the likelihood of delirium in hospitalized patients, said the \pquote{prediction needs to be\ldots\ improved and validated first, which we are actively doing}{P8},
and added a license with a disclaimer and limitation-of-liability clause before sharing the tool with colleagues. 
Caution also extended to participants' own data: having lost work in an earlier project when a code change \pquote{deleted all my work}{P5}, P5 hesitated to use the flashcard app for class and first wanted \pquote{to make sure that the actual software was not buggy, because otherwise, it could totally delete}{P5}.
After a week of use, reliability received the lowest rating among the study-specific measures ($M=3.00$; \cref{fig:perception}).

\mysec{Resisting feature creep (grey-lit: 1,957 posts, 32.21\%, obs).}
As personal software is often created to focus on the needs it serves (\cref{sec:rq1}), creators resisted features that add little value, increase complexity or maintenance burden, or pull the software beyond that fit~\cite{thompson_feature_2005}.
Creators deliberately limited scope,
\glquote{I deliberately scoped it to the current local date only}{G7}, 
removed features they did not use,
\glquote{found myself never using it, and ditched it to keep the interface simple}{G2},
and left out features that would be hard to maintain,
\glquote{Leave out features if they can't be implemented simply, or would require more than minimal server-side processing}{G29}.
Observed participants also restrained their software: P7 told the agent \pquote{Let's not implement editing for now}{P7}, 
and after a week of use P5 removed an AI-added feature they had found \pquote{useless}{P5}.

However, AI coding agents can make this restraint harder because generation is cheap,
\glquote{Scope creep became harder to resist because everything felt cheap}{G18}.
Observed participants faced the same pressure: seven encountered functionality the AI added that they did not want, including redundant or unnecessary interface elements.
Some did not notice such additions: asked whether the agent had added anything unrequested, P6 said
\pquote{Everything has been\ldots\ what I've asked}{P6}, although the software by then contained features the agent had proposed.

\mysec{Context as infrastructure (grey-lit: 1,679 posts, 27.64\%, obs).}
Personal software is built around its creator's situation, so an AI agent can tailor it only as far as it knows that situation: 
the creator's routines, data, devices, and preferences, the people the software serves, and the decisions made about it. 
Most of this context lives only with the creator, and agents do not retain it between sessions, so recording it where agents can retrieve it becomes infrastructure for keeping personal software personal~\cite{jiang_beyond_2026}.
Losing context forces reconstruction,
\glquote{having to re-explain context I had already established}{G4}, and
creators kept project instructions, roadmaps, and plans as
\glquote{scaffolding that makes multi-session development coherent}{G19},
from a Markdown file of project context (\glref{G4}) to a personal context engine that shared history and preferences across its creator's agents (\glref{G6}).

In the observation study, five participants encountered situations in which relevant requirements, assumptions, or context were not shared between themselves and the AI, and five encountered requirements that the AI misunderstood.
Participants often recognized a wrong result once they saw it but not what the agent was missing, so repeated prompting served partly to rebuild that context.

\subsection{RQ3: AI Lowers the Barrier to Building, but Engineering Challenges Remain}
\label{sec:rq3}

RQ3 asks what opportunities and challenges creators encounter when building and
using personal software; it draws on six themes covering 5,218 posts (85.89\%; 2,408 Reddit and 2,810 web posts).
AI coding agents lowered the barrier to building: people without programming
experience built working tools, and all eight observation participants reached
a first working version within four hours. Creators also learned through
making, and sharing could extend a tool's value to others with similar needs.
The challenges lay in the engineering work that remained. Simple-looking
features hid hard technical problems such as data storage and synchronization,
upkeep and external dependencies took ongoing time, and serving others required
revising single-user assumptions and further work.

\mysec{Opportunity: AI lowered entry barrier (grey-lit: 1,872 posts, 30.81\%, obs).}
Creator accounts described AI as lowering the threshold for building software that would otherwise have been out of reach~\cite{myers_past_2000, sarkar_vibe_2025}.
Prompt-based creation lets people build without coding. A creator
replacing the utility sites they repeatedly used explained,
\glquote{Since I don't code, I relied entirely on the prompts}{G3}.
A household meal-planner account described the outcome directly:
\glquote{A non-programmer built an app with Codex}{G33}.
A designer's first app likewise became a working tool used by their partner
to coordinate watering their plants (\glref{G4}).

Observed participants experienced this directly:
seven of eight received useful technical assistance, and
four self-reported limited app-development knowledge or experience.
For example, one participant had considered creating a clinical decision-support tool for months, but \pquote{one of the reasons that we hadn't pursued it was because none of us had ever developed a web app}{P8}.
All eight participants reached a first working version after 40 minutes to under four hours of building, and all six follow-up respondents used their software during the following week; three said they would definitely keep using it.

\mysec{Opportunity: Learning through making (grey-lit: 1,502 posts, 24.72\%, obs).}
Creators can develop technical skills and self-understanding by building
personal software, both while deciding what to build and while using it.
Making a small tool can turn an unfamiliar technology into usable, situated
knowledge. For example, an AI-assisted journal creator gained an explanatory
model of locks and deadlock by asking questions while debugging a frozen save
operation: \glquote{I learned it because I needed to understand why the fix worked, not just that it did}{G9}.
Deciding what to ask the software to do can also reveal what creators want.
A notes-tool creator redesigned tasks and notes around how soon action was
needed, rather than the distinction imposed by the previous apps (\glref{G23}),
and the journal creator found that describing the desired behavior clearly to
an AI was itself a process of deciding what they wanted (\glref{G9}).

In the observation study, participants learned through building and use.
For example, P7 studied the generated code, \pquote{learning what each code means, and why}{P7}.
Use also revealed participants' own routines: after a week, P1 found it
\pquote{easier to see how sleep and exercise affect my mood}{P1}.
Delegation could also limit this learning: \pquote{I'm not really writing the code so I don't know what the AI did differently}{P5}.

\mysec{Opportunity: Value through sharing (grey-lit: 2,766 posts, 45.53\%, obs).}
Sharing personal software can reveal wider needs and extend its value beyond the creator.
A tool initially built for one person or household can address needs shared by peers, communities, or wider audiences, and sharing, feedback, and word of mouth can reveal this broader usefulness and support adoption, reuse, or commercialization.
These opportunities can involve freely shared tools or paid products, without requiring mass-market success.

Creator accounts describe personally useful tools finding value beyond their original audience. 
A meeting-notes creator encountered demand by showing a personal tool to friends:
\glquote{After showing it to a few friends, some asked if it could work for their teams}{G22}.
Those requests revealed wider needs, rather than established team adoption.
A bib generator first used for eleven friends at its creator's birthday run later received \glquote{five payments totalling €65}{G1} from strangers who found it through search. 
A creator's disk-space viewer was adopted by roommates and, after a public release, shared by other users with friends;
the SpaceMonger creator reflected,
\glquote{As it turned out, my problem was everybody else's problem too.}{G40}
Sharing could reveal a common need and bring adoption or payment beyond the
people for whom a tool was initially made.

In the observation study, most participants were willing to share what they had built: five of the six follow-up respondents would share their software, mostly with bounded audiences of people close to them or with the same need, such as friends, family, colleagues, or \pquote{other people who are job searching}{P6}.
Some already saw value beyond themselves: 
P8 had already shared a prototype \pquote{with colleagues, who were excited}{P8}.

\mysec{Challenge: Hidden technical complexity (grey-lit: 1,688 posts, 27.79\%, obs).}
A personal tool can be small and simple to use, but can raise many hard technical problems that its creator did not anticipate and found difficult to resolve,
such as state management, synchronization, and integration.
For example, keeping a reading position and annotations in sync across the Merrilin creator's own devices appeared to be a small addition:
\glquote{These feel like small features, and I had some gumption of how severe they were going to be, but I had no real idea.}{G42}
That addition became three weeks of work on device discovery, pairing, and conflict resolution. 
For a note-taking tool with deliberately limited features, folder support \glquote{took way longer than expected}{G39}, and its creator \glquote{spent the entire day fixing syncing issues}{G39}.

In the observation study, what participants built was small: each tool served one personal purpose with 6 to 27 functionalities. Even so, two participants had difficulty connecting their software to external services, such as sign-in
and databases, maps, or AI models, and three had trouble storing data, finding
it, or keeping it in sync across devices.
Data syncing was the most frequently hidden problem: only two participants realized the
problem before the scenario walkthrough, and only one solved it during the
building phase. P4 found out only during the walkthrough:
\pquote{Close the tab and refresh it, everything resets\ldots\ I think we definitely need a backend}{P4}.
Even after participants realized the problem, fixing it took many rounds of requests.

\mysec{Challenge: Maintenance burden (grey-lit: 1,596 posts, 26.27\%, obs).}
Keeping personal software working takes ongoing time from its creator, both for the software and its data and for the platforms, APIs, and AI models it depends on. 
These dependencies reduce the work a personal project requires, but they bring prices, usage limits, and policies that the creator cannot control or negotiate. 
Whether the creator makes time for this upkeep can decide whether the software stays in use.
In creator accounts, creators take on every fix,
\glquote{bugs and features are my problem}{G30}, along with the care of their data,
\glquote{Personal data lives outside git, so backup is critical}{G10}.
This work competes with the rest of their lives, 
\glquote{The expensive input was my evenings}{G1}; one creator shut down a
baby-name service after their child was born because keeping it running
\glquote{wouldn't be sustainable at this time in my life}{G27}.
Dependencies also add upkeep that creators do not choose. 
For example, a family messaging app had to be rebuilt after its previous service shut down (\glref{G28}), and a task-app creator found that
\glquote{the sovereignty I had over my Windows Phone app was limited by the collapse of the entire platform}{G30}.

In the observation study, similarly, three participants were unsure during building how they would modify or maintain the software, and upkeep took time as soon as use began.
Only three of the six follow-up respondents changed their software during the week; the others cited limited time or dependence on additional data. 
For those who did, repair could crowd out use, \pquote{I spend more time fixing bugs or adjusting appearance than actually using the app}{P5},
and P5 \pquote{used up my daily gemini tokens trying to fix one bug}{P5}.

\mysec{Challenge: Sharing imposes additional work (grey-lit: 2,536 posts, 41.74\%, obs).}
Although sharing can extend the value of personal software (\emph{Value through sharing}), serving others takes more than a tool that works for its creator. 
It requires revising assumptions made for one user and adding infrastructure, support, and release work, which can make broader distribution difficult for a solo creator.
A creator warned prospective users of a photo manager built around a desktop workflow, 
\glquote{you'll likely have needs it won't meet}{G14}, 
and supporting mobile access was out of reach: \glquote{That would be a giant rewrite}{G14}.
Turning a personal grocery-slot script into a public service meant \glquote{maintaining a queue of searches from different users}{G41}, and a plan to put it online over a weekend became fourteen weekends of work.

In the observation study, all five participants who would share their software estimated between a few hours and more than a week of further work before others could use it, and none rated its commercial potential above moderate. 
Some of this work concerned reliability: P5 would share the flashcard app with the general public only
\pquote{once I catch all those annoying bugs and feature annoyances}{P5}.
Serving others also unsettled single-user assumptions: P4 had assumed that \pquote{everyone will be logging on their own phone}{P4} until the agent suggested that roommates might use one another's devices.

\section{Discussion}
\label{sec:discussion}

We discuss how our findings relate to prior accounts of end-user programming,
homegrown tools, and vibe coding, and what they imply for supporting personal
software.

\mypara{Optimism despite code-and-fix.}
Compared with traditional development processes, personal software development with AI agents resembled an Agile process~\cite{beck2001agileprinciples} if the creator and the agent are viewed as a team: with the agent as the developer and the creator as the customer, requirements changed as working software revealed them~\cite{beck2001agileprinciples}, and 89 of the 131 functionalities (68\%) emerged during building (\cref{sec:rq2}).
In practice, however, the process resembled code-and-fix~\cite{boehm1988spiral}: participants prompted the agent to fix whatever they found, focusing on functionality and leaving design, testing, and maintenance for later.
Boehm~\cite{boehm1988spiral} attributed three difficulties to code-and-fix: poor fit to users' needs, code that becomes so poorly structured after a number of fixes that later fixes are very expensive, and costly fixes due to poor preparation for testing and modification.
Being their own users, participants caught poor fit early, spending 47\% of building time testing or reading (\cref{fig:timeline}).
Prior work shows that AI-assisted development is not exempt from the other two: projects that adopted Cursor saw lasting increases in code complexity~\cite{he_speed_2026}.
Our participants nonetheless stayed optimistic (\pquote{I can just add it}{P2}): after a week of using and building their software, they rated their confidence in modifying it $M=4.67$ of 5 (\cref{fig:perception}), and the five willing to share expected 1--3 hours to more than a week (median 1--2 days) of further work before others could use it.
Early signs of these difficulties did appear, but they did not change this impression: fixes frequently took several attempts, \glquote{Each fix regressed something else}{G17}, and P5 \pquote{used up my daily gemini tokens trying to fix one bug}{P5}, yet still rated their confidence in modifying the software 4 of 5.
This optimism may arise because creators with little programming experience cannot anticipate the technical complexity behind their software (\emph{Hidden technical complexity}); in the grey literature, one creator's plan to put a tool online over a weekend took fourteen weekends (\glref{G41}).
It may also be because most participants did not intend to share broadly, which reduced their sense of responsibility for other users and the need for cross-platform testing. Moreover, building personal software in our study cost participants little or nothing, which may have made them more willing to accept the risk of future maintenance costs.

\mypara{Defaults can override needs.}
The agentic coding tools' defaults shaped personal software, sometimes against creators' needs.
For example, Google AI Studio added unrequested Gemini features.
For P5, Google AI Studio added article summaries to P2's journal even though P2 preferred to read the original articles.
Structure also followed the tool rather than the need: Google AI Studio gave all four of its apps the same React/Vite scaffold, whereas Claude, an open-ended tool, initially produced P3's and P4's software each as a single HTML file, even when the software had complex functionality to support.
The tools applied these defaults regardless of each creator's needs and requests, and creators often lacked the programming expertise to recognize or override them.
Future tools should follow creators' needs more closely, or help creators recognize these defaults and request changes explicitly. For example, a tool could ask whether the software should be a single file or have a backend.

\mypara{Weaker and stronger barriers.}
In a previous canonical paper, Ko, Myers, and Aung~\cite{ko2004six} identified six learning barriers in end-user programming systems: design, selection, coordination, use, understanding, and information.
In our observation study, AI coding agents weakened some of these barriers, making them easier to overcome, and strengthened others, making them harder.
Selection and use weakened the most: the agent chose frameworks and libraries and wrote the code that used them, and participants edited code by hand in only 0.2\% of building intervals (\cref{fig:timeline}).
Coordination weakened inside a well-designed product such as Google AI Studio, where the agent could connect the different components using pre-configured guidance, but stayed hard when the software needed to connect with external services: two participants had difficulty connecting it to external services, and three had trouble storing data or keeping it in sync across devices (\emph{Hidden technical complexity}).
Design also weakened, as participants could start without knowing everything they wanted and decide while building: 89 of the 131 functionalities (68\%) emerged during building (\cref{sec:rq2}).
In contrast, the understanding and information barriers grew stronger, because participants were trying to fix features that they may not understand the mechanism behind.
All eight saw their software behave in ways they did not expect, and six had to ask the agent several times for the same change, sometimes not knowing what to ask: \pquote{I don't even know what to tell it or give it}{P5}.
Checking was also hard: three struggled to tell what the agent did, and two thought a task was done when it was not (\emph{Shift to validation}).
End-user programming has focused on helping people write code.
With AI coding agents, end-user got software without actually writing code but often could not tell why it misbehaved or how to fix it; support from tool and education material should seek to help them understand and fix what the agent built (\cref{sec:implications}).

\mypara{Homegrown tools outside the company.} Smith et al.~\cite{smith_build_2015} studied homegrown tools that Microsoft developers built on their own initiative.
Developers in their study built tools to work around existing tools that did not fit their environment, to save time and reduce the pain of repetitive tasks, and to help teammates. These reasons match the \emph{Personal fit} and \emph{Community fit} that our creators described (\cref{sec:rq1}).
Smith et al.~\cite{smith_build_2015} suggested that developers tolerate less of the pain of repetitive tasks because they can automate these tasks; AI coding agents may further extend that ability, and that lower threshold, to people who cannot program (\emph{AI lowered entry barrier}).
Those Microsoft developers were also kept the tools within their team and mentioned the polishing needed before sharing as a barrier, much like the further work our study expected before others could use their software (\emph{Sharing imposes additional work}).
The main difference between homegrown tools inside the company and outside the company is the support around the creator.
At Microsoft, managers granted time, team cultures encouraged building, and colleagues took over maintenance of tools that proved useful~\cite{smith_build_2015}.
By contrast, personal software creators often work without such organizational support, so maintaining and sharing the software falls on the creator and the agent (\emph{Maintenance burden}; \emph{Sharing imposes additional work}).

\mypara{The safeguards that are absent.} 
Chou et al.~\cite{chou2026building} found that vibe coders used undo, version control, small changes, automated tests, and linters to rein in the agent's stochastic changes.
We did not observe our participants using version control, automated tests, or linters.
They verified 93 of the 131 functionalities, but mainly by checking that the visible result looked right (\cref{sec:rq2}).
When something broke, repair went through the same channel that had produced the problem: another prompt to the agent. Only P8 wanted to undo a change, and even P8 planned to ask the agent rather than use the restore feature AI Studio provides.
One possible reason that our participants did not use these safeguards is programming experience: in the study by Chou et al.~\cite{chou2026building}, the safeguards came mostly from professional programmers, and tests and linting came only from them.
Another possible reason, specific to undo, is that undoing is harder when an agent manages the project: in a code editor, undoing a change takes a single command, whereas one agent response can change many things at once.
If people without programming backgrounds are to build personal software, which the grey-literature corpus shows is already happening at scale, safeguards cannot depend on creators importing them from experience, and undoing an agent's change should be as simple as undoing an edit.

\mypara{Ownership through involvement.}
Creators tied ownership to understanding or steering their software, and that sense could fade with continued use.
Some creator accounts gained ownership by clearly understanding where their data lived and what the code did, \glquote{I know exactly what happens to my data}{G21}, while others lost it when they only act on approving the agent's changes, \glquote{I started to feel like I was renting the code rather than owning it}{G18}.

Our participants did both.
During building, P7 directed the agent and read parts of the generated code, and said \pquote{It's mine. I know exactly what it does}{P7}, similar to the novices in the study by Louie et al.~\cite{louie_novice-ai_2020}, who felt more ownership of their music when they could steer the AI.
P2 had felt attached at the end of building, having suggested features and adjusted the app to their taste.
A week later, P2 rated ownership 1, the lowest among participants ($M=3.33$; \cref{fig:perception}) and stopped using the app, planning a \pquote{new app developed by myself}{} because \pquote{I want an app that I have sense of ownership over}{P2}.
Steering the agent during building may thus not sustain ownership, and software that fits its creator may still be abandoned once ownership fades. Future work should quantify the relationship between ownership and involvement and explore how to sustain ownership, for example by helping creators understand the code and data behind their software.

\section{Implications}
\label{sec:implications}
We discuss six implications for supporting people who build personal software. 
\mypara{Learn the creator's situation before building.}
Personal software is built around one person's situation, yet the agents in our study sometimes built without asking any clarification and filled the gaps with generic assumptions.
Five participants found that the agent misunderstood a requirement, and five found that they and the agent lacked shared context, leaving them unsure what to ask or what the agent meant (\emph{Context as infrastructure}). P4's household coordinator was meant only for P4's own household, but the agent built onboarding for new households because it \pquote{considered that this app may be for\ldots\ commercial use}{P4}.
P8's agent did not ask about the coefficients of the delirium prediction model and had \pquote{come up with its own weights, which are totally made up}{P8}.
The scenario walkthrough, a step we added after the pilot sessions, surfaced such gaps at little cost: right after participants declared their software done, they wrote at least three situations from their own lives in which they would use it and then tried each one on the software (\cref{sec:observation-method}).
At the end of this process, participants found something to change in 20 of their 28 scenarios (\cref{sec:rq2}).
AI coding tools could therefore ask creators for such scenarios before building, including who will use the software and on which devices, instead of assuming them.

\mypara{Show the assumptions the agent made.}
Not every detail can be settled before building and the agent filled some details the creator left open with its own assumption.
Only the creator could tell whether an assumption fit, and the same agent could be right and wrong for the same person.
While building a clinical tool, P8 encountered both a correct and a wrong assumption: the agent \pquote{anticipated what I didn't say, that I wanted to be HIPAA compliant}{P8}, but the agent also turned P8's request to save data as a CSV file into a download button, which P8 did not want and asked it to remove this feature afterwards.
Participants often learned of such assumptions only when they ran into the result: seven encountered functionality the agent added that they did not want (\emph{Resisting feature creep}), sometimes without knowing its purpose, \pquote{I never asked for this, and they just added that. I don't know why they did that}{P5}.
After undoing one such change, P8 resolved to \pquote{ask it for more information before asking it to do anything}{P8}.
AI coding tools could make this the default by stating the assumptions behind a change before applying it, such as who will use the software, where its data is stored, and what goes beyond the request, so that creators can confirm or correct each one.

\mypara{Going back is missing.}
One agent response can change many parts of the software at once, so creators need a dependable way back to a version that worked.
Our participants instead repaired problems the same way they had caused them, by prompting again; to undo a change in AI Studio, P8 planned to \pquote{ask it to undo that}{P8}; the other participants did not undo at all.
Undoing through a prompt is less direct than undoing an edit, and prompting was not reliable: six participants had to ask several times for the same change (\emph{Context as infrastructure}), as P6 put it, \pquote{I kept telling it to fix it multiple times, and nothing really changed}{P6}.
Integrating git into AI coding tools could help, but it needs certain interfaces to be usable by people without programming experience. 
AI coding tools therefore need an interface that makes undoing an agent's change as easy and reliable as pressing Ctrl+Z in a text editor.

\mypara{Better verification for end users.}
Validating the result took most participants' building time: 47\% of the annotated intervals showed them testing the software or reading the agent's output (\cref{fig:timeline}).
One reason is that the tools offered little help beyond what participants could see on the screen (\emph{Shift to validation}): when the agent fixed errors in the code behind it, P8 noted, \pquote{this is the stuff that's all on the backend that I can't see}{P8}.
Each change also invokes additional rechecks across other components, because a fix in one place could break another: in P5's flashcard app, the agent would \pquote{break the LaTeX and then fix the text}{P5}, so once the LaTeX finally worked, P5 did not want \pquote{to touch it, because otherwise the LaTeX might ruin itself}{P5}.
The WYSIWYT methodology showed that end-user programmers could test spreadsheets effectively when testing was embedded in the artifact they were using~\cite{rothermel_wysiwyt_2000}.
Platforms for building personal software should also embed testing and verification in an easy way for end users, such as by showing which parts of the software are affected by a change and what are some checking suggestions.

\mypara{Let creators take control over their code and data.}
Some creators built personal software to regain control from commercial products, yet building on hosted AI tools created new dependencies (\emph{Maintenance burden}).
While building, P5 and P8 both ran out of AI Studio's quota.
Moving the code elsewhere was not easy either; when P8 was using AI Studio's GitHub connection, saving the code to a repository took about ten minutes.
And when a service disappears, the software may have to be rebuilt, as happened to a family messaging app (\glref{G28}).
Platforms that host AI-built software could therefore let creators export their code and data in standard formats and state alternative services end-user could rely on.
In the EU, the Data Act requires cloud providers, including SaaS providers, to facilitate the transfer of customers' exportable data and digital assets~\cite{eu_data_act_2023}. However, it does not necessarily ensure that software remains functional after migration, particularly when it depends on proprietary services. Meaningful portability therefore should extend beyond a one-time export.

\mypara{Study ownership and upkeep beyond the first week.}
We followed participants for one week of use, and two findings call for longer studies.
First, ownership varied widely: the six participants who answered the survey a week later rated it from 1 to 5 (\cref{fig:perception}), for different reasons.
P8 found that \pquote{the nicer it looked, the more I wanted the sense of ownership over it}{P8}; P3, impressed that the agent built a prototype \pquote{in such little time}{P3}, felt no attachment to it; and P2's attachment at the end of building had faded a week later (\cref{sec:discussion}).
Why ownership forms or fades, and whether tools that involve creators in the agent's decisions can sustain it, needs further investigation.
Second, participants remained optimistic that they could keep modifying their software ($M=4.67$ of 5; \cref{fig:perception}), although code-and-fix and prior studies of AI-assisted development predict that fixes grow costly as software ages~\cite{boehm1988spiral, he_speed_2026}.
Longitudinal studies could test whether this confidence holds and learn when creators maintain, rebuild, or abandon their software.

\section{Threats to Validity}
\label{sec:threats-to-validity}

\mypara{Grey-literature corpus.} Our corpus over-represents creators who chose to write publicly about their software, and Reddit  subreddits skew toward technically engaged users.
People who abandoned personal software quietly are underrepresented.
Exa's semantic search may also miss accounts that describe the same practice in different terms, although we derived 75 descriptions until no new terms emerged.
Creator accounts are retrospective and self-selected, so we treat them as evidence of the concerns creators articulate, not as a representative sample of outcomes.
We still used grey literature because public posts are the most accessible proxy for privately built software at scale, and we followed established guidelines~\cite{garousi2019guidelines} in our analysis to best capture the corpus of creators' experiences and concerns.

\mypara{LLM-assisted screening and coding.} An LLM screened grey-literature documents for relevance and generated their initial codes.
We validated screening against human labels ($\kappa=0.75$) and kept theme development with the research team, but screening may have excluded accounts written in unfamiliar or domain-specific language, and the three coding perspectives were roles of one model rather than independent coders.
Manual analysis was infeasible at this scale, so we followed prior LLM-assisted approaches~\cite{agarwal2026opinions,ahmed2025can,qiao_thematic-lm_2025}.
LLMs may hallucinate paragraph identifiers or misattribute quotations to posts, preventing us from tracing back to the original sources from codes. When generating post counts, we excluded 6,234 of 173,700 quotation links (3.59\%) that had unknown identifiers. These exclusions affected only the reported post counts; the code-to-theme assignments and thematic structure were unchanged.
In the observation study, we compared LLM and human annotations for functionality planning and evolution on a held-out participant ($\kappa=0.701$) and for on-screen activity on two participants ($\kappa=0.753$).
While these values indicate substantial agreement, individual labels may still contain errors.

\mypara{Observation sample and setting.} Given the length of each participant's involvement, from a building phase of up to several hours to a week of use, and the exploratory nature of our study, we believe eight participants are sufficient for this exploratory study. However, we do not claim to estimate prevalence; 
the percentages we report describe our sessions only. 
While we tried to maximize diversity, recruitment through social media, forums, and campus networks favored people interested in AI tools, and most participants were 18 to 24. 
Our findings may therefore only partially applicable to older creators, creators unfamiliar with technology, or creators from underrepresented groups, who might have different difficulties, such as accessibility issues. 
Think-aloud and the researcher's presence may have changed how participants worked, and four participants stopped building when asked whether they were done or when the session had to end, so we do not treat time to a minimal viable product as a finding.
Participants also used different AI coding tools, and we cannot isolate tool effects and some difficulties may be tool-specific.

\mypara{Functionality grouping.} Our functionality-level results depend on how we grouped requests into functionalities. 
An LLM rated the similarity of each participant's requests, and we clustered them rather than grouping them by hand to make the process reproducible.
A different grouping could merge or split functionalities and change the reported shares. However, it is unlikely to change the overall finding that most functionalities emerged during building or the evolution result, since we used the same grouping for all analyses; any merges or splits would thus apply consistently across participants.

\mypara{Duration and tool familiarity.} Six of the eight participants completed the seven-day follow-up. 
Seven days can show initial adoption or abandonment but not the long-term maintenance that creator accounts describe, so our claims about continued use are limited to this window. 
We chose seven days to limit participant burden~\cite{bolger2003diary} and rely on creator accounts for longer-term use.
Several participants were also using AI coding tools for the first time, so how they prompted, built, and verified their software, as well as the difficulties and perceptions they reported, may partly reflect their unfamiliarity with the tools rather than stable practice for personal software creation.
However, creators in the grey literature, who were not limited to first-time users, reported similar patterns, such as effort shifting to validation (\emph{Shift to validation}), unwanted AI additions (\emph{Resisting feature creep}), and ongoing upkeep (\emph{Maintenance burden}). The mixed-methods design therefore cross-validates the findings and ensure a more complete picture of personal software creation process.

\section{Conclusion}
\label{sec:conclusion}

We studied personal software, the software people build primarily for their own
use, through a grey-literature review of creator accounts and an observation
study of people building software for their own needs with AI coding agents.
Across both studies, we found characteristics and motivations that distinguish
personal software from professionally developed software and from the practices
studied in end-user programming and vibe coding research: its value lies in
fitting a particular life rather than serving many users, its requirements
emerge as creators react to what they build, and its creator is often its only
developer, tester, and maintainer. AI coding agents change who can
build personal software and where its engineering work falls. We
therefore see \textit{personal software engineering} as an area that deserves
study in its own right, and we call for research on how to support this work,
including quality standards suited to software built for one person and
proportional to its stakes, tools that capture creators' intent and make agent
decisions visible, and longitudinal studies of how personal software is used
and maintained over time. AI coding agents bring the decades-old vision of a
personal, programmable computer closer, but realizing it depends on this
engineering work.

\section{Data Availability}
\label{sec:data-availability}
The study protocol, a supplemental appendix with additional
justifications, and the grey-literature sources
are available at \url{https://anonymous.4open.science/r/supp-pse-EFF3/}.
We will make the package publicly available upon acceptance.

\bibliographystyle{ACM-Reference-Format}
\bibliography{references}

\appendix
\crefalias{section}{appendix}
\newpage
\section{Grey-literature review materials}
\label{sec:appendix-grey-literature}

\subsection{Web Search Details}
\label{sec:appendix-web-search}

\paragraph{Prefix selection.}
The research team selected the seven prefixes based on the research questions
and a preliminary inspection of at least 20 results for each candidate prefix, checking
that each prefix produced meaningful queries when combined with the
descriptions. The prefixes, reproduced verbatim, were
\texttt{account of creating}, \texttt{account of making},
\texttt{experience creating}, \texttt{experience making},
\texttt{why I built}, \texttt{How i built}, and \texttt{What i built}.

\paragraph{Descriptions of personal software.}
To derive the 75 descriptions, each researcher collected at least 20 conceptual
posts about personal software, from which we extracted definitions and related
terms and iteratively searched for additional terms until no new ones emerged.
We then merged synonymous phrases and removed ambiguous, technology- or
time-specific, and outcome-specific wording. \cref{tab:exa-descriptions} lists
the descriptions verbatim. Each query consisted of a prefix, a space, and a
description, and requested up to 50 results.

\paragraph{Recovering missing text.}
Based on a preliminary manual review of 400 results, we developed five
regular-expression checks for potential paywalls and used HTTP response codes
to flag inaccessible pages. We manually reviewed flagged pages and supplemented
their text where possible. Of 312 flagged results, we recovered content for 308;
for the remaining four, we retained only publicly accessible material.

\begingroup
\footnotesize
\setlength{\tabcolsep}{3pt}
\renewcommand{\arraystretch}{1.12}
\begin{longtable}{@{}p{0.05\textwidth}p{\dimexpr0.95\textwidth-6pt\relax}@{}}
\caption{Descriptions of personal software combined with the seven prefixes in
the web search, reproduced verbatim.}
\label{tab:exa-descriptions}\\
\toprule
\textbf{No.} & \textbf{Description} \\
\midrule
\endfirsthead
\toprule
\textbf{No.} & \textbf{Description} \\
\midrule
\endhead
\midrule
\multicolumn{2}{r}{Continued on next page}\\
\endfoot
\bottomrule
\endlastfoot
1 & Software built not for a market segment, a team, or a company, but for a person. Sometimes just one person. \\
2 & Software for hobbies, households, personal data, odd routines, niche workflows, and temporary needs. \\
3 & Software created when someone noticed a friction in their own life, had the means to do some software about it, and built the exact thing they needed. \\
4 & single-use or custom applications that are adapt to you, not the other way around, by you with AI. \\
5 & Softwares that aren't factory-made and it is made for someone, not for everyone \\
6 & Software that wouldn't make sense to most people but is exactly right for the people who use it \\
7 & softwares that don't need to be sold \\
8 & software built in creating something that has no audience beyond yourself and a few friends. \\
9 & Any software developed to fit a need unique to a person, system, or team. \\
10 & Software written for an audience of one. \\
11 & An application or a script built to solve a specific problem for a specific person, with no immediate intention of scaling, monetizing, or even sharing. \\
12 & A software that is not commercial software (built for scale), boutique software (tailor-made for companies and organizations) and open source software (structured for collaboration). \\
13 & micro-tools made by ``anyone'' for themselves and the people around them \\
14 & Software made by people who understand the problem more deeply than the underlying technology to solve it \\
15 & Software emerges from a personal relationship (a friend's request, a colleague's frustration) or it establishes one. \\
16 & software where anyone can adapt to their needs with minimal friction. \\
17 & small tweaks to existing software to creating new tools that work well in coordination with existing ones. \\
18 & software designed in and for a particular social situation or context. \\
19 & software built with AI, free and fully under one person's control \\
20 & software made with love and care, serve your specific needs, private, without financial pressure. \\
21 & software solves local problems for local people. \\
22 & Software that user have full agency and power over their own creations \\
23 & highly specialized tools for their unique workflows, hobbies, and problems. \\
24 & Software doesn't need to scale to millions of users, handle edge cases beyond its creator's needs, or maintain backward compatibility across versions. \\
25 & software built by agentic systems for one person \\
26 & one-time scripts designed for a specific use case without considering future needs \\
27 & software created without external customers in mind \\
28 & Software Not intended for sale, sharing, or scaling, Not meant for GitHub, enterprise, or general audiences, to fit a specific problem that a specific person has and that's it. \\
29 & software developed by someone for their own use (or, perhaps, for the use of a handful of people), fitting their particular needs and preferences, running and providing data access across devices. \\
30 & software that are narrow enough to be executed by a single person \\
31 & software which should be operable in local-only or self-hosted setups with minimal effort \\
32 & tools built for specific moments, needs, or contexts \\
33 & software is purpose-built to solve my own niche problems. It's not designed to scale or solve everyone's problem and that's okay! \\
34 & small applications that solve a personal problem or task, without the need to always make it an app that people can buy or download. \\
35 & software doesn't need product-market fit---it just needs to fit you. \\
36 & small, personal applications we build for ourselves, tailored to our specific needs. \\
37 & tiny, personal tools that solve individual problems in highly specific ways. \\
38 & Software that are not professionally-built, robust systems, like operating systems, or commercial applications built for broad audiences. \\
39 & Software that are not as a product, but as a gift for ourselves and the people in our life \\
40 & Software built for a small group, for a specific purpose, without needing to cater to the whims of venture funding or justify itself with engagement metrics \\
41 & A tool just for you, just for this time \\
42 & Software where the boundary between users and creators becomes increasingly fluid, which is enabled by advances in large language models and AI-assisted development tools. \\
43 & Software created by making yourself the user you're most eager to please. \\
44 & software created by writing code for yourself without any external customers in mind. \\
45 & software that is lightweight, functional, and designed from the start to be made yours. \\
46 & software that is small, and quirky, and targeted at a single-digit audience. \\
47 & software that is small things that worked only for me, and `works on my machine' was basically all I needed. \\
48 & personalized, community-driven, and locally controlled applications \\
49 & applications designed for specific social contexts and small groups, rather than large-scale, generic solutions. \\
50 & small applications written entirely for my own benefit \\
51 & Software applications intentionally designed for the needs, practices, and local contexts of particular communities of practice, rather than a broader, more generic set of ``users''. Personalised, localised, often created by the community who's using it. \\
52 & Software which is built and personalized to a single person or organization. The features, interface, capabilities, and usage are all created for the specific organization to use. \\
53 & Software where the minimum viable customer base is you. \\
54 & tools that perfectly match our individual needs and workflows. \\
55 & An app can be a home-cooked meal \\
56 & Software that is ruthlessly simple; we love it; no one else will ever use it. \\
57 & Software won't change unless we want it to change. There will be no sudden redesign, no flood of ads, no pivot to chase a userbase inscrutable to us. It might go away at some point, but that will be our decision \\
58 & Software that is small, simple, and makes me super happy. \\
59 & custom software you build for yourself, to solve your own specific problems, with no compromises and no audience. \\
60 & Software created by senior+ engineers, the ones who know how to architect distributed systems and run incident reviews to spend their evenings building something that will only ever have one user - themselves \\
61 & Software with properties: Ownership --- your data, your domain, your files; Durability --- open formats outlast proprietary ones, Plain text outlasts everything; Sovereignty --- no one else decides when your tools change, disappear, or start showing ads; Right-sizing --- not everything needs to scale; Agency --- building your own tools, tending your own garden, cooking your own meals, etc. \\
62 & Software built for yourself with No compromises, no bloat, no unused features, No what if someone wants to..., No extension frameworks, No plugin architecture. If you ever need it to do something else, you can extend it because you wrote it. \\
63 & simple and small software built by individuals with agentic AI to solve a personal problem. \\
64 & Software that you build for yourself, you optimize for one person's needs, and you can cut every corner you want because you are the only one who bears the consequences \\
65 & remarkable and super niche software applications by and for consumers. \\
66 & software for any nuanced or custom need \\
67 & Software build exactly what you needed, for yourself, for your workflow, for your specific situation. \\
68 & Software that is small, specific, or too personal to monetize \\
69 & Software that is built for exactly your workflow, your data, your one weird edge case, which compromises for no one and nothing ever leaves your own boundary. \\
70 & Software whose creation imposes a cognitive load small enough to be carried by one person---the system should fit in one head. \\
71 & Software whose user experience should satisfy the needs and preferences of its designer, as opposed to those of some generic user \\
72 & software doesn't need to solve the problem of its economic sustainability. \\
73 & customized, personal tools for yourself---tools where you're the only user. \\
74 & Software that only exists because you needed it, shaped entirely by your constraints, and maintained by asking questions instead of writing configs, works better than anything you could buy because it was built for exactly one team: yours \\
75 & applications created by users, tailored to their specific needs using AI-assisted development tools. \\
\end{longtable}
\endgroup

\subsection{Reddit Filtering}
\label{sec:appendix-reddit-filtering}

To focus on project accounts in five AI-related subreddits, we retained posts with flairs
containing ``built with claude'' in ClaudeAI, ``project'' in ChatGPTCoding, and
``showcase'' in GitHubCopilot, lovable, and GoogleAIStudio.

\subsection{Relevance Classifier}
\label{sec:appendix-relevance}

Human raters labeled 404 posts (297 web, 107 Reddit). Two raters independently
labeled 102 web posts ($\kappa=0.7397$), after which one rater labeled the
remaining 195 web and 107 Reddit posts. We used 200 labeled web posts to refine
the prompt and evaluated the classifier on held-out sets of 97 web posts and
107 Reddit posts.

The constrained JSON schema returned the implemented artifact, its original
purpose, a brief rationale, and a binary relevance decision. Three calls per
post provided a minimal majority-voting procedure while limiting
computational cost. An input safeguard limited each call to the first 200,000
characters; 14 web posts and three Reddit posts exceeded this limit.

\cref{tab:relevance-metrics} reports the classifier's performance by source.

\begin{table}[htbp]
\centering
\footnotesize
\caption{Relevance-classifier performance, treating relevance as the positive
class.}
\label{tab:relevance-metrics}
\begin{tabular}{@{}lrrrrr@{}}
\toprule
Evaluation set & $n$ & Agreement & Precision & Recall & $\kappa$ \\
\midrule
Exa & 97 & 86.60\% & 90.91\% & 81.63\% & 0.7322 \\
Reddit & 107 & 88.79\% & 84.78\% & 88.64\% & 0.7700 \\
Combined & 204 & 87.75\% & 87.78\% & 84.95\% & 0.7523 \\
\bottomrule
\end{tabular}
\end{table}

Combined F1 was 0.8634.

\paragraph{Prompt.}
Each call sent the following system prompt.
\begin{quote}
\footnotesize\setlength{\parindent}{0pt}\setlength{\parskip}{0.5em}
Decide whether the page is relevant evidence about creating personal software. Set keep=true only for actual implemented software made initially for its maker's own need or a bounded known person, group, situation, or workflow.

First identify the implemented artifact and its earliest purpose. One-off scripts and tools for self, family, one specific team, client, class, hackathon, or community count. A broader essay can count when its maker's actual implemented tool or scripts are substantial evidence, even if used to draw lessons. Reject conceptual or hypothetical discussion with no actual instance, and thematic guides or surveys where unrelated projects merely illustrate advice. Explicitly reported real projects from one named class, team, community, or place count even if extraction omits linked project names.

Origin controls the decision. Later release, open source, users, sale, SaaS, or broad adoption do not erase an explicit personal/bounded origin. But solo authorship, side-project or repository status, local/self-hosted design, technical detail, or a narrow feature alone do not prove personal origin. Reject software initially made for income, a market/customer/audience class, research participants, public infrastructure, or a business's clients, members, students, or users generally. One named client or team is bounded; a customer category is not.

A dedicated maker project/repository page can count with unstated origin when it documents a concrete end-user app or tool and gives no market-first evidence. This default excludes company sales pages, libraries, frameworks, servers/platforms, and explicit market-first projects. An aggregate page counts only if it substantially documents one implemented bounded system beyond listing products.

Reject unavailable/error pages, pure documentation, unrelated pages, and non-software. Ignore trailing boilerplate, navigation, comments, and related links. Think silently. Report the artifact evidence and earliest-purpose evidence, then give a brief reason and one final keep decision.
\end{quote}
The user message contained the post's URL and text.
\begin{quote}
\footnotesize
\begin{verbatim}
URL: {url}
<POST>
{text}
</POST>
\end{verbatim}
\end{quote}

\subsection{Corpus Sampling and Segmentation}

Reddit sampling used equal allocation across the 10 subreddits. Strata smaller
than their allocation were included in full, and unused places were
redistributed equally among the remaining strata. This yielded 190 posts from
BuildWithClaude, 314 from GitHubCopilot, 32 from GoogleAIStudio, and 364 from
each of the other seven subreddits. A fixed random seed (42) made the sample
reproducible. The analytic corpus covered all 1,928 web domains in the retained
web corpus.

Units exceeding 30,000 characters were split further to remain below Excel's
cell-size limit, producing 259,939 units. The loader interpreted eight
paragraphs containing only \texttt{None} as missing values, leaving 259,931
coding units. The dataset retained source URLs, post identifiers, unit
order, and character offsets.

\subsection{Coding Pipeline Parameters}
\label{sec:appendix-coding-pipeline}

The batch sizes (20 units per coder request, 100 units per aggregation batch)
balanced request size against the overhead of repeated aggregation and review
calls. Retrieving a limited set of 10 neighbors kept review requests manageable
as the codebook grew. Reviews proceeded in batch order so that each incorporated
the preceding codebook updates. The cap of 20 quotations per code bounded
storage and downstream context.

\paragraph{Coder identities.}
Each of the three coders received one of the following identities. Thematic-LM
adds the identity to the coder's system prompt and asks the coder to let that
background inform how it interprets and labels the data.
\begin{quote}
\footnotesize\setlength{\parindent}{0pt}\setlength{\parskip}{0.5em}
You are a software engineer with years of industry experience building and maintaining applications, who also builds small tools for your own use and regularly works with AI coding agents.

You are a researcher in human-computer interaction and software engineering who studies how developers and end users build software, including with AI programming tools.

You work in a non-technical field, have little or no formal programming experience, and have built a few small applications for your own life, judging them by whether they do what you need.
\end{quote}

\paragraph{Maximum theme count.}
To capture the range of creators' needs, practices, and challenges, we compared
maximum theme counts from 10 to 100 in the theme-generation prompt, assessing
overlap, granularity, and redundancy, and selected 50.

\paragraph{Theme-generation prompt.}
We sent the following prompt to \texttt{gpt-6-luna}, followed by the full
codebook with one code per line and the number of its supporting quotations in
brackets.
\begin{quote}
\footnotesize\setlength{\parindent}{0pt}\setlength{\parskip}{0.5em}
You are doing reflexive thematic analysis of a large corpus to answer these research questions:\\
- What characterizes personal software and the needs it serves?\\
- How do people create and evolve personal software?\\
- What opportunities and challenges do creators encounter when building and using personal software?

Below is the full codebook from an earlier coding pass: every code, with the number of supporting quotes in brackets (a rough salience signal). Read it as a whole.

Propose 5 to 50 candidate THEMES that answer the research questions.

What a theme is, and is NOT:\\
- A theme is a PATTERN OF SHARED MEANING organized around a central concept -{}- an analytic CLAIM that says something interesting about the phenomenon. Write each title as a claim/argument, NOT a topic label like ``AI tooling''.\\
- A theme CUTS ACROSS many codes; it is not a single code renamed, and not a category of similar codes.

For each theme give: a compact, representative 3-8 words abbreviated title; a claim-style title; a one-to-two sentence claim; the central organizing concept; and a brief note of which parts of the codebook it draws on. Use only what is supported by the codebook.
\end{quote}

\newpage
\subsection{Codebook}
\label{sec:appendix-codebook}

\cref{tab:codebook} lists the five theme groups and 17 themes with their
definitions. Group names follow \cref{fig:theme}. \emph{Validation scales with
consequence} belongs to two groups and has a definition for each.

\begingroup
\footnotesize
\setlength{\tabcolsep}{3pt}
\renewcommand{\arraystretch}{1.12}
\begin{longtable}{@{}>{\raggedright\arraybackslash}p{0.2\textwidth}p{\dimexpr0.8\textwidth-6pt\relax}@{}}
\caption{Codebook of theme groups and themes developed in the grey-literature review.}
\label{tab:codebook}\\
\toprule
\textbf{Theme} & \textbf{Definition} \\
\midrule
\endfirsthead
\toprule
\textbf{Theme} & \textbf{Definition} \\
\midrule
\endhead
\midrule
\multicolumn{2}{r}{Continued on next page}\\
\endfoot
\bottomrule
\endlastfoot
\multicolumn{2}{@{}l}{\textbf{Characteristics}} \\
Personal fit & \emph{Personal software creates value by fitting specific lives and circumstances.} Personal software often starts from routines, constraints, relationships, or tastes that general-purpose tools do not serve well. Its value can come from fitting those circumstances closely, even when the result is not broadly applicable to others. \\
Community fit & \emph{Software can be valuable when it fits a specific community rather than a broad market.} Tools grounded in a shared language, place, practice, or community can support coordination, preserve local knowledge, and strengthen collective activity. Their value may depend more on situated trust and relationships than on broad reach, while adoption, inclusion, moderation, and sustainability still require attention. \\
Bounded scope & \emph{The scope of personal software can remain bounded by the problem it serves.} Personal software can stay small because it is bounded by the need it serves, and meeting that need can be enough for its creator to regard it as complete. \\
Controlled data and privacy & \emph{Personal software can provide greater data control while creating privacy and design responsibilities.} Users may value being able to keep, inspect, move, and govern their data and tools through local-first storage, open formats, configurable behavior, and self-hosting. When sensitive information is involved, this control also requires choices around processing, minimization, permissions, sharing, and security. These approaches can strengthen agency and trust while adding setup, maintenance, or technical complexity. \\
\addlinespace
\multicolumn{2}{@{}l}{\textbf{Creation}} \\
Iteration through use/test & \emph{Real use reveals gaps and drives repeated refinement.} A working demo or successful local test does not guarantee dependable use. As builders and users rely on software in practice, mismatched assumptions, missing data, compatibility problems, confusing flows, and edge cases become visible, prompting changes to behavior, scope, and design. \\
Resisting feature creep & \emph{Deliberate restraint can preserve usefulness as new features become easier to add.} As adding functionality, integrations, and engagement mechanisms becomes easier, builders can expand beyond the need that made a tool useful. Deliberate omissions and incremental additions can preserve a clear purpose, reduce maintenance and user burden, and avoid unnecessary software bloat. \\
Learning through making & \emph{Building personal software can develop both technical skill and self-understanding.} Creating and using a tool can teach makers about technologies as well as their own needs, habits, and abilities. AI can make experimentation more accessible, but durable understanding still develops through examining outcomes, troubleshooting, and deciding what matters. \\
Hidden technical complexity & \emph{Simple personal tools can conceal substantial engineering work.} A lightweight interface or one-command workflow may depend on careful work in data modeling, security, synchronization, compatibility, error handling, and recovery. User-facing simplicity therefore does not necessarily imply a technically simple system. \\
\addlinespace
\multicolumn{2}{@{}l}{\textbf{Continuous usage}} \\
Intrinsic motivation & \emph{Noncommercial rewards can motivate creators to keep building.} Learning, enjoyment, creative expression, agency, and meaningful contribution can sustain personal software projects even when few people use them and no revenue follows. Curiosity, play, milestones, public sharing, and supportive peers can reinforce that motivation, while comparison, hostile criticism, and pressure to monetize can weaken it. \\
Easy integration & \emph{Personal tools are easier to keep using when they fit existing workflows and services.} People often already rely on calendars, messaging, documents, editors, and specialized platforms. Interfaces and adapters that connect these systems can preserve familiar habits and reduce the burden of adopting another silo, although fragile integrations and platform changes can threaten continued use. \\
Maintenance burden & \emph{Ongoing upkeep can determine whether personal software remains useful over time.} Dependencies, platforms, data, user needs, and external services continue to change after the initial build. Greater use or reliance can turn support, updates, backups, operations, and responses to changes in APIs, pricing, access, policies, or technical behavior into recurring work that consumes time and attention. Builders may reduce these risks through portability, local alternatives, self-hosting, or from-scratch solutions, but doing so can introduce additional implementation and maintenance work. Maintenance therefore becomes a central condition of longevity. \\
\addlinespace
\multicolumn{2}{@{}l}{\textbf{AI usage}} \\
Lowered entry barrier & \emph{AI lowers the barrier to turning more ideas into working software.} Natural-language prompting and coding agents let more people prototype ideas, explore unfamiliar technologies, and produce working applications with less prior programming experience. This expands both what feels feasible to build and who can participate in software creation. \\
Shift to validation & \emph{AI shifts human effort from implementation toward validation and oversight.} As AI increases the speed and volume of implementation, people still need to specify goals, provide context, review changes, test behavior, and decide what belongs. Human attention therefore moves toward direction, verification, and decision-making, and multiple agents can intensify that demand. \\
Context as infrastructure & \emph{Effective AI-assisted work depends on preserving and retrieving relevant context.} Personal workflows and agent tasks rely on histories, preferences, decisions, project state, and domain knowledge that can otherwise be lost between interactions. Capturing and retrieving the right context can reduce repeated work and improve specificity, while stale, excessive, or poorly organized context can mislead. \\
Validation scales with consequence & \emph{More consequential AI actions require stronger validation, human review, and limits on autonomy.} Low-consequence assistance may tolerate lightweight checks, while AI that affects money, health, privacy, or essential operations requires more rigorous validation. Human review, reversible actions, scoped permissions, and explicit limits on automation become more important as the consequences of error increase. \\
\addlinespace
\multicolumn{2}{@{}l}{\textbf{Expansion}} \\
Value through sharing & \emph{Sharing personal software can reveal wider needs and extend its value beyond the creator.} Tools initially built for one person or household can address needs shared by peers, communities, or wider audiences. Sharing, feedback, and word of mouth can reveal this broader usefulness and support adoption, reuse, or commercialization. These opportunities can involve freely shared tools or paid products, without requiring mass-market success. \\
Sharing imposes additional work & \emph{Reaching and serving others requires more than a tool that works for its creator.} Finding and retaining users requires discovery, promotion, explanation, onboarding, navigating platform rules, and earning trust. Serving those users can also require revising creator-specific assumptions and adding interface design, privacy protections, permissions, reliability, infrastructure, and support. Differences in users, data, environments, and concurrent use can demand architectural or governance changes. This additional product, operational, and engineering work can make broader distribution difficult or unsustainable for a solo creator. \\
Validation scales with consequence & \emph{Broader reliance can require stronger validation and safeguards.} A personal tool may tolerate lightweight testing and rough edges when only its creator relies on it, but broader adoption can increase both the number of people affected by failures and the importance of dependable behavior. Validation, reversible actions, permissions, and operational safeguards therefore become more important as reliance grows. \\
\end{longtable}
\endgroup

\section{Grey-Literature Sources Referenced in the Paper}
\label{sec:appendix-grey-literature-sources}

Post IDs identify the original web and Reddit posts referenced in the findings and
discussion. The full list of posts is in \texttt{grey-lit-valid-posts.json} in the replication package.
Links and quoted passages were checked on 2 October 2026.

\begingroup
\footnotesize
\setlength{\tabcolsep}{3pt}
\renewcommand{\arraystretch}{1.12}
\begin{longtable}{@{}p{0.06\textwidth}p{0.40\textwidth}p{\dimexpr0.54\textwidth-12pt\relax}@{}}
\caption{Post IDs, original titles, and source links for the referenced grey literature.}
\label{tab:quoted-grey-literature}\\
\toprule
\textbf{Post ID} & \textbf{Title} & \textbf{Original link} \\
\midrule
\endfirsthead
\toprule
\textbf{Post ID} & \textbf{Title} & \textbf{Original link} \\
\midrule
\endhead
\midrule
\multicolumn{3}{r}{Continued on next page}\\
\endfoot
\bottomrule
\endlastfoot
\hypertarget{gl:G1}{\glref{G1}} & I built a race bib generator for my own birthday run & \url{https://bib-gen.com/blog/i-built-a-race-bib-generator-for-my-own-birthday-run} \\
\addlinespace
\hypertarget{gl:G2}{\glref{G2}} & Hibiscus.txt & \url{https://memoryofsnow.com/notes/hibiscus/} \\
\addlinespace
\hypertarget{gl:G3}{\glref{G3}} & Built an all-in-one Utility App with zero coding experience (3 weeks, \textasciitilde{}150 credits) & \url{https://www.reddit.com/r/lovable/comments/1qr1ixa/built_an_allinone_utility_app_with_zero_coding/} \\
\addlinespace
\hypertarget{gl:G4}{\glref{G4}} & A designer's first attempt at building with Claude Code & \url{https://medium.com/design-bootcamp/a-designers-first-attempt-at-building-with-claude-code-571686dfa17f} \\
\addlinespace
\hypertarget{gl:G5}{\glref{G5}} & Home cooked apps & \url{https://olickel.com/my-home-cooked-meals} \\
\addlinespace
\hypertarget{gl:G6}{\glref{G6}} & ctx: The Personal Context Engine Every One of My Agents Shares & \url{https://zackproser.com/blog/ctx-personal-context-engine} \\
\addlinespace
\hypertarget{gl:G7}{\glref{G7}} & Building Dream Journal: a local-first journaling app built for privacy and low friction & \url{https://www.sebasab.com/blog/building-dream-journal} \\
\addlinespace
\hypertarget{gl:G8}{\glref{G8}} & Bubbyway, the Household App I Wrote for the Two of Us & \url{https://lee-it.co.uk/software/2026/05/13/bubbyway-a-household-app-for-two.html} \\
\addlinespace
\hypertarget{gl:G9}{\glref{G9}} & Seven Months of Vibe Coding: How I Built a Privacy-First Journal App with an AI Pair Programmer & \url{https://www.kennethlacroix.me/blog/seven-months-of-vibe-coding/} \\
\addlinespace
\hypertarget{gl:G10}{\glref{G10}} & Building a Personal AI Agent: Architecture and Lessons from LifeOS & \url{https://pkarnal.com/blog/building-my-ai-agent} \\
\addlinespace
\hypertarget{gl:G11}{\glref{G11}} & Build Yourself Small Tools & \url{https://v5.chriskrycho.com/journal/build-yourself-small-tools/} \\
\addlinespace
\hypertarget{gl:G12}{\glref{G12}} & Designing Personal Software & \url{https://alexsci.com/blog/personal-apps/} \\
\addlinespace
\hypertarget{gl:G13}{\glref{G13}} & I Built a Personal Finance Tracker That Fits in a Single File & \url{https://medium.com/@mudunuruchandu/i-built-a-personal-finance-tracker-that-fits-in-a-single-file-e5c2615d0404} \\
\addlinespace
\hypertarget{gl:G14}{\glref{G14}} & Software for one & \url{https://zerotrickpony.com/articles/software-for-one/} \\
\addlinespace
\hypertarget{gl:G15}{\glref{G15}} & Just completed a 632 line UI to Text script & \url{https://www.reddit.com/r/ChatGPTCoding/comments/187pbs5/just_completed_a_632_line_ui_to_text_script/} \\
\addlinespace
\hypertarget{gl:G16}{\glref{G16}} & How I Vibe-Coded a Village Tool Lending Library & \url{https://medium.com/@michaelakolackova/how-i-vibe-coded-a-village-tool-lending-library-cb254c29680b} \\
\addlinespace
\hypertarget{gl:G17}{\glref{G17}} & The Last Carpenter & \url{https://www.kuril.in/blog/the-last-carpenter/} \\
\addlinespace
\hypertarget{gl:G18}{\glref{G18}} & I Built a CRM Twice with AI & \url{https://cstanton.net/posts/i-built-a-crm-twice-with-ai} \\
\addlinespace
\hypertarget{gl:G19}{\glref{G19}} & Agentic Engineering: Building Without Writing & \url{https://dehora.net/journal/2026/3/agentic-engineering-building-without-writing} \\
\addlinespace
\hypertarget{gl:G20}{\glref{G20}} & Building a Gratitude App I Wanted to Use & \url{https://www.retiredpdx.com/p/building-a-gratitude-app-i-wanted} \\
\addlinespace
\hypertarget{gl:G21}{\glref{G21}} & Journal v1.0: A Simple Go Project That Grew Up & \url{https://jamiehurst.co.uk/2026-04-29_journal-v1} \\
\addlinespace
\hypertarget{gl:G22}{\glref{G22}} & I Built an AI Meeting Notes Tool in 4 Hours for \$0.32/hr & \url{https://yu-wenhao.com/en/blog/ai-meeting-notes/} \\
\addlinespace
\hypertarget{gl:G23}{\glref{G23}} & Build tools around workflows, not workflows around tools & \url{https://thesephist.com/posts/tools/} \\
\addlinespace
\hypertarget{gl:G24}{\glref{G24}} & The joy of cooking (an app) & \url{https://huonw.github.io/blog/2021/08/home-cooked-app/} \\
\addlinespace
\hypertarget{gl:G25}{\glref{G25}} & Khaganate: a suite of personal productivity software & \url{https://iafisher.com/blog/2022/02/khaganate} \\
\addlinespace
\hypertarget{gl:G26}{\glref{G26}} & Go create silly, small programs & \url{https://jakemccrary.com/blog/2020/10/03/go-create-silly-small-programs/} \\
\addlinespace
\hypertarget{gl:G27}{\glref{G27}} & How Namestork evolved from an idea to a tiny service over four years & \url{https://tatey.com/2022/01/04/how-namestork-evolved-from-an-idea-to-a-tiny-service-over-four-years/} \\
\addlinespace
\hypertarget{gl:G28}{\glref{G28}} & An app can be a home-cooked meal & \url{https://www.robinsloan.com/notes/home-cooked-app/} \\
\addlinespace
\hypertarget{gl:G29}{\glref{G29}} & Designing Low Upkeep Software & \url{https://www.jefftk.com/p/designing-low-upkeep-software} \\
\addlinespace
\hypertarget{gl:G30}{\glref{G30}} & App Sovereignty & \url{https://ezhart.com/posts/i-wrote-another-to-do-app} \\
\addlinespace
\hypertarget{gl:G31}{\glref{G31}} & Clobsidian in Detail: Cross-Source Personal Infrastructure & \url{https://simon.podhajsky.net/blog/2026/clobsidian-in-detail/} \\
\addlinespace
\hypertarget{gl:G32}{\glref{G32}} & Build Your Own CLI Tool With AI in an Afternoon & \url{https://ricardolara.dev/blog/build-your-own-cli/} \\
\addlinespace
\hypertarget{gl:G33}{\glref{G33}} & My new retirement hobby: I just built my first app, in less than a week & \url{https://www.beginningwithi.com/2026/03/03/retirement-hobby-built-my-first-app-in-under-a-week/} \\
\addlinespace
\hypertarget{gl:G34}{\glref{G34}} & The Race to a Beta Launch: The Trials and Tribulations of Piggy Back & \url{https://avandamiri.com/2009/07/16/piggy-back-launch-trials-and-tribulations.html} \\
\addlinespace
\hypertarget{gl:G35}{\glref{G35}} & How I Built Cute Desk App in 32 Days with an AI Coding Partner & \url{https://cutedesk.app/blog/how-i-built-cute-desk-app-in-32-days.html} \\
\addlinespace
\hypertarget{gl:G36}{\glref{G36}} & Not Everything Is About The Money & \url{https://blog.ewancroft.uk/3mm2rj3ejg22x} \\
\addlinespace
\hypertarget{gl:G37}{\glref{G37}} & Odyssey Projects & \url{https://blog.jread.com/posts/odyssey-projects/} \\
\addlinespace
\hypertarget{gl:G38}{\glref{G38}} & Why I still build apps for myself & \url{https://blog.kulman.sk/why-i-still-build-ios-apps/} \\
\addlinespace
\hypertarget{gl:G39}{\glref{G39}} & building WebNotes - the whole story so far\ldots{} & \url{https://aetos-devlogs.hashnode.dev/building-webnotes-the-whole-story-so-far} \\
\addlinespace
\hypertarget{gl:G40}{\glref{G40}} & The SpaceMonger 1.x Post-Mortem & \url{https://www.werkema.com/programming/the-spacemonger-1-x-post-mortem/} \\
\addlinespace
\hypertarget{gl:G41}{\glref{G41}} & How I Built and Launched Find a Delivery & \url{https://blog.karannavani.com/how-i-built-and-launched-find-a-delivery} \\
\addlinespace
\hypertarget{gl:G42}{\glref{G42}} & What Was I Reading Last? In Three Not-So-Easy Pieces & \url{https://tech.stonecharioteer.com/posts/2026/merrilin-local-sync/} \\
\addlinespace
\hypertarget{gl:G43}{\glref{G43}} & Why I built my own static site generator & \url{https://yakko.dev/blog/teeny} \\
\addlinespace
\hypertarget{gl:G44}{\glref{G44}} & Apps without ambition & \url{https://booplet.com/blog/personal-apps/} \\
\addlinespace
\end{longtable}
\endgroup

\section{Observation Study Materials}
\label{sec:appendix-observation}

\subsection{Procedure Details}

\paragraph{Idea exploration.}
Participants first described the software they wanted to create or a concrete
situation in which they wished they had software to help them. We then asked
about its purpose, their reasons for wanting it, and how often the situation
arose. If participants mentioned existing software, we asked about difficulties
or friction with it. We also asked them to categorize the software they wanted,
describe how they expected to use it, and name it, so that they had a concrete
concept of what they would build.

\paragraph{Choice of Google AI Studio.}
AI Studio is a well-known platform for the Gemini API, and its
build-an-app feature also offers a low-friction path from creation to continued
personal use and sharing. Its free tier lets participants keep using and
modifying their apps without paying for ongoing development, configuring
hosting, or provisioning an API quota, which reduces barriers to sustained use.
It also imposes relatively few architectural assumptions, so backend services
and other integrations can be introduced when needed rather than having the
development environment strongly shape what participants build. We introduced
its basic functionality to participants without prior agentic-coding
experience; the introduction script is in the supplemental material.

\paragraph{No time limit.}
Because ideas can change and evolve both within and outside the building
process, we imposed no time limit and no limit on the number or length of
breaks. We reminded participants of the time two hours into the
session, but they were free to continue as long as they wished.

\paragraph{Think-aloud prompting.}
We used the time the agent spent generating to prompt think-aloud, asking about
the goal participants were pursuing and what they expected the agent to
produce. Roughly every one to two hours, when a prompt was likely to take more
than 300 seconds, we asked additional questions about decisions they had made,
whether their ideas had changed, and what they planned next. Participants could
tell us when they preferred not to be interrupted, and we then refrained from
asking questions until they were ready.

\paragraph{Devices and contexts of use.}
During the pilot sessions, many participants did not consider using their
software across devices. We therefore asked participants to specify the device
or context of use whenever their usage scenarios did not.

\paragraph{Survey tools.}
The Session~1 evaluation survey used Google Forms. The Session~2 survey used
QuestionPro for its question branching, logic, and data-prefilling features. 
\paragraph{Compensation rationale.}
Participants were paid for their time at the minute level. To attract
participants motivated to build personal software rather than to earn money,
we kept rates modest, and we paid a lower rate for building because
participants spent much of that time waiting for the agent. The higher rate for
the follow-up interview acknowledged the additional effort of returning to the
study.

\paragraph{Pilot observations.}
Beyond the need for usage scenarios and cross-device testing, the pilot
sessions suggested several patterns of human--AI decision making during
building, which led us to capture participants' decision processes and the
evolution of their requirements.

\subsection{Functionality Grouping}
\label{sec:appendix-grouping}

A locally hosted LLM (\texttt{qwen3.6-35b-a3b-q6}) split each prompt into
sub-requests and rated the similarity of every pair of a participant's
sub-requests from 0 to 1, judging whether they addressed the same feature area.
We clustered the sub-requests with agglomerative hierarchical clustering on
distances derived from these ratings, searching five similarity-to-distance
transformations, four linkage methods (average, complete, weighted, and single),
and every number of clusters, and kept the partition with the highest
silhouette score (0.47--0.73 across participants). The model then named each
cluster. After removing three clusters of automatically generated requests such
as ``Continue,'' 131 functionalities remained (6--27 per participant), covering
396 sub-requests from 206 prompts.

\subsection{LLM-Assisted Coding Details}
\label{sec:appendix-llm-coding}

\cref{tab:llm-coding} lists the model, number of runs, and human--LLM
agreement for the functionality, on-screen action, and verification tasks.

\begin{table}[htbp]
\centering
\footnotesize
\setlength{\tabcolsep}{3pt}
\caption{LLM-assisted coding and human--LLM agreement (Cohen's $\kappa$; Jaccard similarity for verification).}
\label{tab:llm-coding}
\begin{tabular}{@{}l>{\raggedright\arraybackslash}p{2.6cm}c>{\raggedright\arraybackslash}p{3.4cm}@{}}
\toprule
Task & Model & Runs & Agreement \\
\midrule
Functionality planning and evolution & \texttt{qwen3.6-35b-a3b-q6}, temp.\ 0 & 3 & 0.832 (validation, P3); 0.701 (test, P8) \\
On-screen actions & \texttt{qwen3.6-35b-a3b-q6}, temp.\ 0 & --- & 0.753 (two-second frames) \\
Verification & \texttt{qwen3.6-35b-a3b-q6}, temp.\ 0 & --- & 0.83 (P1 and P3) \\
\bottomrule
\end{tabular}
\end{table}

For the action and verification annotations, we ran \texttt{qwen3.6-35b-a3b-q6} with model thinking disabled. We evaluated the verification annotations on P1 and P3 because they used different development environments (Google AI Studio and Claude), which could affect how verification evidence appeared in the recordings. Given this variation in development environments, we did not hold out any participants for the on-screen action and verification annotations. We then used the LLM-generated annotations for the remaining six participants to analyze functionality planning and evolution and on-screen activity.

\subsection{Codebooks}
\label{sec:appendix-obs-codebooks}

Tables~\ref{tab:obs-codebook-motivation}--\ref{tab:obs-codebook-actions} give the
codebooks of the observation study. 
In the functionality-evolution
codebook, the first request about a functionality can be coded \emph{Abandoned} when it
asks to remove that functionality, most likely because the AI implemented it and
the participant was not satisfied. The on-screen action codes adapt the CUPS
taxonomy of \citet{mozannar2024reading}, with testing and reading merged into
one code.

\begingroup
\footnotesize
\setlength{\tabcolsep}{3pt}
\renewcommand{\arraystretch}{1.12}
\begin{longtable}{@{}p{0.05\textwidth}>{\raggedright\arraybackslash}p{0.27\textwidth}p{\dimexpr0.68\textwidth-12pt\relax}@{}}
\caption{Motivation codebook.}
\label{tab:obs-codebook-motivation}\\
\toprule
\textbf{ID} & \textbf{Code} & \textbf{Definition} \\
\midrule
\endfirsthead
\toprule
\textbf{ID} & \textbf{Code} & \textbf{Definition} \\
\midrule
\endhead
\midrule
\multicolumn{3}{r}{Continued on next page}\\
\endfoot
\bottomrule
\endlastfoot
M01 & Promote their work (connect with domain expertise) & The creator aims to translate their research or domain expertise into software that makes their work accessible and useful to others. \\
M02 & Better Take Care of Others & The creator aims to improve the experience or care of others in their community. \\
M03 & Personal life/routine & The creator wants the software to support a task, situation, or routine in their own everyday life. \\
M04 & Learn AI-assisted development & Learn how to use AI tools to build software or develop relevant technical skills. \\
M05 & Existing Tool Complexity/friction & The creator seeks a simpler or more usable solution because existing tools, models, or workflows are too complex for their intended use, or no existing tool exists. \\
M06 & Funding/budget constraints & Limited or unavailable funding for software development/usage. \\
M08 & Existing tool friction > Using multiple software tools for one task & The creator uses several software tools to complete one task or workflow and wants to bring that work together. \\
M09 & Existing tool friction > Hard to communicate and change existing tool & The creator has difficulty communicating desired changes to those responsible for an existing tool or obtaining those changes. \\
M10 & Existing tool friction > Limited awareness or unwillingness to find existing tools & The creator is unaware of potentially relevant existing tools or has not searched for them, or is unwilling to do so. \\
\end{longtable}
\endgroup

\begingroup
\footnotesize
\setlength{\tabcolsep}{3pt}
\renewcommand{\arraystretch}{1.12}
\begin{longtable}{@{}p{0.05\textwidth}>{\raggedright\arraybackslash}p{0.27\textwidth}p{\dimexpr0.68\textwidth-12pt\relax}@{}}
\caption{Opportunity and difficulty codebook, by category.}
\label{tab:obs-codebook-oppdiff}\\
\toprule
\textbf{ID} & \textbf{Code} & \textbf{Definition} \\
\midrule
\endfirsthead
\toprule
\textbf{ID} & \textbf{Code} & \textbf{Definition} \\
\midrule
\endhead
\midrule
\multicolumn{3}{r}{Continued on next page}\\
\endfoot
\bottomrule
\endlastfoot
\multicolumn{3}{@{}l}{\textbf{Creator--platform interaction challenges}} \\*
C01 & Limited app-development knowledge or experience & The creator lacks the technical knowledge or practical experience needed to develop, evaluate, or manage the application. \\
C02 & Requirements misunderstanding or insufficient clarification & The AI misinterprets the creator's intended requirements, or insufficient clarification leaves the requested behavior or constraints ambiguous. \\
C03 & Uncertainty about future modification and maintenance & The creator is unsure whether or how the generated software can be updated, modified, or maintained as needs change. \\
C04 & Limited visibility into AI actions and progress & The creator cannot determine what the AI is doing, whether it is progressing, or what actions it has completed. \\
C05 & Repeated prompting to obtain the requested change & The creator must repeat or reformulate a request because the AI has not successfully made the intended change. \\
C06 & External-service integration difficulties & The creator encounters obstacles connecting the application or development platform to external services, accounts, or infrastructure. \\
C07 & Difficulty removing obsolete files & The creator struggles to remove unnecessary files or determine whether deleting them will disrupt the application. \\
C08 & Dissatisfaction with the quality of LLM-generated code & The creator expresses dissatisfaction with the quality of code generated by the LLM. \\
C09 & Limited ability to manage the generated software & The creator lacks the available controls or a feeling of lacking controls needed to modify, configure, or otherwise manage the generated software as intended. \\
C10 & Incorrect/Inaccurate AI-provided information & The AI supplies incorrect or misleading factual content, explanations, instructions, or claims about the software and its actions. \\
C11 & Lack of shared information between creator and AI & The creator and AI lack shared context, reference points, requirements, or assumptions needed to coordinate their work, including information available to only one party. \\
C12 & High debugging and troubleshooting effort & Identifying, diagnosing, and resolving problems requires substantial time, effort, or repeated intervention from the creator. \\
C13 & Diminished sense of ownership/authorship & The creator expresses reduced or limited ownership of, or identification as the author of, the software because of AI involvement. \\
C14 & False assumption of AI task completion & The participant believes that the AI has completed a requested task, often because the AI or platform indicates that it has, but later testing, inspection, or use reveals that the task was incomplete, unsuccessful, or not actually performed. \\
C15 & Errors in LLM-generated code & The generated code contains an error that prevents intended execution or produces incorrect behavior. \\
\multicolumn{3}{@{}l}{\textbf{Difficulties using and managing the generated software}} \\*
D1 & Data storage and access difficulties & The creator encounters problems determining or ensuring where data is stored, whether it persists, or how it can be retrieved and accessed appropriately. \\
D2 & Difficulty locating generated files & The creator cannot find or identify a file that the AI generated or claimed to have generated. \\
D3 & Unwanted AI-added functionality & The AI introduces unwanted functionality or the generated interface contains duplicated or unnecessarily repeated controls, information, or interaction elements. \\
D4 & Unexpectedly slow AI execution & The AI takes longer than the creator expects to carry out a requested action or produce a result. \\
D5 & Version-control difficulties & The creator encounters problems setting up and managing repositories, commits, pushes, or other version-control operations. \\
D6 & Mismatch between expected and delivered behavior & The creator explicitly identifies a difference between how they expected the generated software to behave and how it actually behaves. \\
D7 & Difficulty interpreting generated outputs & The creator finds the software's displayed results or information difficult to understand because their meaning, presentation, or context is unclear. \\
\multicolumn{3}{@{}l}{\textbf{Opportunities of using an AI-assisted development environment}} \\*
O1 & Technical assistance from AI & The creator expects or experiences useful AI assistance with technical implementation, problem-solving, or development decisions. \\
O2 & Mockups/illustrations build confidence & Seeing mockups, illustrations, or previews increases the creator's confidence that the intended software can be built or will be usable. \\
O3 & General positive appraisal & The creator expresses satisfaction, enthusiasm, or a favorable overall evaluation without identifying a more specific benefit. \\
O4 & AI expands the creator's design options/ideas & The AI introduces useful or welcomed features, approaches, or design possibilities that the creator had not previously considered. \\
O5 & Flexible ways to modify and sync the generated software in one platform & Being able to modify the software through different methods, such as conversational requests, or direct code editing in one platform. \\
O6 & Helpful examples and explanations & AI-provided examples or explanations help the creator understand a concept, implementation, or possible course of action. \\
O7 & Easy external-service integration through built-in support & Built-in platform capabilities make connecting the software to external services straightforward or easier than the creator expected. \\
O8 & Sense of ownership/authorship & The creator expresses a sense that the software is their creation or reflects their contribution despite AI involvement. \\
O9 & Appreciation of platform tools for working with the software & The creator positively evaluates platform tools that support inspecting, editing, previewing, testing, or publishing the generated software. \\
\end{longtable}
\endgroup

\begingroup
\footnotesize
\setlength{\tabcolsep}{3pt}
\renewcommand{\arraystretch}{1.12}
\begin{longtable}{@{}>{\raggedright\arraybackslash}p{0.2\textwidth}p{\dimexpr0.8\textwidth-6pt\relax}@{}}
\caption{Functionality-evolution codebook.}
\label{tab:obs-codebook-evolution}\\
\toprule
\textbf{Code} & \textbf{Definition} \\
\midrule
\endfirsthead
\toprule
\textbf{Code} & \textbf{Definition} \\
\midrule
\endhead
\midrule
\multicolumn{2}{r}{Continued on next page}\\
\endfoot
\bottomrule
\endlastfoot
Introduced & Requests a new functionality or deliverable that has not previously been requested or implemented. \\*
Expanded & Adds a distinct capability to an existing functionality. \\*
Refined & Adds detail to or modifies an existing functionality or requirement without adding a distinct capability or correcting a defect. \\*
Narrowed & Reduces a functionality's scope or permitted use while retaining its core capability. \\*
Repaired & Requests correction of an error or failure to meet an existing requirement. \\*
Abandoned & Explicitly removes or drops an existing or previously requested functionality. \\*
Restated & Repeats an earlier request for the same functionality. \\*
Explanation & Requests explanations, instructions, locations, or recommendations without requesting implementation. \\*
Confirmation & Requests confirmation that a functionality, change, or result exists or works as intended. \\
\end{longtable}
\endgroup

\begingroup
\footnotesize
\setlength{\tabcolsep}{3pt}
\renewcommand{\arraystretch}{1.12}
\begin{longtable}{@{}>{\raggedright\arraybackslash}p{0.2\textwidth}p{\dimexpr0.8\textwidth-6pt\relax}@{}}
\caption{On-screen action codebook.}
\label{tab:obs-codebook-actions}\\
\toprule
\textbf{Code} & \textbf{Definition} \\
\midrule
\endfirsthead
\toprule
\textbf{Code} & \textbf{Definition} \\
\midrule
\endhead
\midrule
\multicolumn{2}{r}{Continued on next page}\\
\endfoot
\bottomrule
\endlastfoot
\texttt{testing\_reading} & Operating or looking at the running app (clicking, entering data, scrolling or navigating its pages, opening its dialogs, running its commands), or reading what the agent wrote: its reply, changelog, code or approval request. Operating the app counts even while the agent works; looking and reading count only when it is not working. \\*
\texttt{editing\_code} & Typing in a code editor: characters appear or disappear. Scrolling or reading code is not editing, nor is code the agent writes or shows for approval. \\*
\texttt{prompting} & The caret is in the agent's input (chat box or terminal prompt), typing or paused mid-draft, until it is sent. \\*
\texttt{configuring} & Operating the tool's own settings: sharing, billing, model, API keys, sign-in, deployment. The app's own settings are testing. \\*
\texttt{other\_window} & Neither the tool nor the app is on screen. \\*
\texttt{waiting} & The agent is working (a live timer such as ``Running for 42s'', a spinner or ``\ldots{}ing'' status, output still streaming) and the person is not operating the app, editing code, prompting, configuring, or in another window: reading or watching its output while it works is waiting. \\*
\texttt{no\_screen} & The camera-off placeholder: no screen is shared. \\
\end{longtable}
\endgroup

\subsection{Functionality Planning Prompts}
\label{sec:appendix-prompt-planning}

Planning is coded from the transcript before the participant's first prompt to
the agent, which includes idea exploration. For each participant line there, the
model lists the functionality groups that the line plans. A group is planned
when at least one line plans it. The model is \texttt{qwen3.6-35b-a3b-q6}
(Qwen3.6-35B-A3B, 6-bit quantization), run locally with llama.cpp at
temperature 0 with thinking off, and with no other sampling setting and no
output-length limit. It returns JSON constrained by a strict schema, and three
runs are combined by majority vote.

\paragraph{System message.}\mbox{}

\begin{lstlisting}[style=prompt]
You are coding transcripts from a qualitative study of people who build software for their own use with an AI agent.

How a session runs: first the researcher interviews the participant about the software they would like to create (the problem behind it, its purpose, what it should look like and do, how they would use it). Then the participant sets up an AI tool and starts building by sending prompts to the AI agent while thinking aloud. The transcript was produced by speech recognition, so wording can be garbled.

The participant's prompts were later split into sub-requests, and the sub-requests were grouped into functionality groups. You get the groups, each listed with its sub-requests; a group stands for the functionality its sub-requests ask for. Judge a group by its sub-requests, not by its name. Some groups only ask the agent to fix, explain or check something that was already built; such a group stands for the specific behaviour its sub-requests want fixed, explained or checked.

Your task: code the transcript before the participant's first prompt, sentence by sentence. For each PARTICIPANT line, list the functionality groups that the line plans. Read every line in its context: the researcher's question it answers and the participant's neighbouring lines.

A line plans a group when the participant expresses that they want, intend or expect their software to do what at least one of the group's sub-requests asks for, whether the line uses the same words as the sub-requests or everyday words for the same behaviour. This includes describing the functionality, how they would use it, or a problem (with existing tools, or with how they manage now) that they want their software to solve, as long as the group's sub-requests ask for that same functionality; a general aim or benefit that the group's functionality would be only one possible way to reach does not plan it. It includes functionality they would like but present as optional or for later, but not functionality they say they do not want or need.

It must be the same functionality, not merely a related topic:
- Wanting to make the software at all, what it is about in general, the AI tool or account used for the study, and going along with the study procedure do not plan any group, even a group whose sub-requests only ask the agent to build the software as a whole or to carry out a step of the study procedure.
- The kind of product it will be, or the device or platform it will run on, plans a group only when that group's sub-requests change the software itself to be that kind of product or to work there (for example its layout or input), not a group whose sub-requests only ask the agent to produce the software or to help the participant get it running.
- A general wish about the software's qualities (how it should feel or look overall) plans a group only when the participant ties it to the specific functionality that group's sub-requests ask for.
- A group that only asks the agent to fix, explain or check something is planned only when the participant wants that specific behaviour, not merely the feature it belongs to.

A short answer plans a group when it accepts, confirms or chooses functionality of that group that the question it answers describes, including accepting it as optional or for later; a short answer that declines or rejects that functionality, or says the participant does not know whether they want it, does not.

A line can plan several groups or none.
\end{lstlisting}

\paragraph{User message.}\mbox{}

\begin{lstlisting}[style=prompt]
Functionality groups of this participant's software (each with the sub-requests the participant later sent to the AI agent):

G1 <group name>
    - <sub-request>
    - <sub-request>
G2 <group name>
    - <sub-request>
…

Transcript before the participant's first prompt to the AI agent:

[<line ID>] <h:mm:ss> RESEARCHER: <text>
[<line ID>] <h:mm:ss> PARTICIPANT: <text>
…

Code these PARTICIPANT lines: <the line IDs, separated by commas>.
For each of them give the list of functionality groups the line plans (an empty list when it plans none).
\end{lstlisting}

\subsection{Functionality Evolution Prompts}
\label{sec:appendix-prompt-evolution}

Each sub-request gets one code from the evolution codebook.
The codes are \emph{Introduced}, \emph{Expanded}, \emph{Refined}, \emph{Narrowed},
\emph{Repaired}, \emph{Abandoned}, \emph{Restated}, \emph{Explanation}, and
\emph{Confirmation}. The prompt uses \emph{Information} and \emph{Verification}
for the last two codes, respectively. The model and settings are those of the planning prompts:
\texttt{qwen3.6-35b-a3b-q6} at temperature 0 with thinking off, returning JSON
constrained by a strict schema. Three runs are combined by majority vote.

\paragraph{System message.}
The codebook and its note are taken from the evolution codebook.

\begin{lstlisting}[style=prompt]
You are coding prompts from a qualitative study of people who build software for their own use with an AI agent.

How a session runs: first the researcher interviews the participant about the software they would like to create. Then the participant builds it by sending prompts to the AI agent while thinking aloud; the researcher asks what each prompt is for. The transcript was produced by speech recognition, so wording can be garbled. Each prompt was split into sub-requests, and the sub-requests were grouped into functionality groups.

Your task: give every sub-request of the prompt being coded one code that describes how it evolves the software's functionality, relative to what came before it. What came before is everything requested in earlier prompts, everything the sub-requests listed before it in the same prompt request, and anything the agent is known to have built (the agent may build things nobody asked for). Code what the sub-request asks for; use the full prompt and the transcript (what the participant says about the prompt and about the software) to understand what they meant.

The groups tell you which sub-requests concern the same functionality, but a group's first sub-request is not automatically Introduced. Group names were written afterwards to summarise all of a group's sub-requests over the whole session; code each sub-request from what it asks for, not from the words in its group's name.

Codebook:
- Introduced: Requests a new functionality or deliverable that has not previously been requested or implemented.
- Expanded: Adds a distinct capability to an existing functionality.
- Refined: Adds detail to or modifies an existing functionality or requirement without adding a distinct capability or correcting a defect.
- Narrowed: Reduces a functionality’s scope or permitted use while retaining its core capability.
- Repaired: Requests correction of an error or failure to meet an existing requirement.
- Abandoned: Explicitly removes or drops an existing or previously requested functionality.
- Restated: Repeats an earlier request for the same functionality.
- Information: Requests explanations, instructions, locations, or recommendations without requesting implementation.
- Verification: Requests confirmation that a functionality, change, or result exists or works as intended.

Note from the codebook: The first request can be abandon when user requesting removal in the first request of one functionality (most likely ai implemented and human not satisfied)

How to tell close codes apart:
- Introduced vs Expanded vs Refined. Introduced is a functionality or deliverable nobody has requested or built yet. Expanded adds a distinct capability, something new the software can do, to a functionality that already exists or was already requested. Refined adds detail to or changes how an existing or already requested functionality or requirement looks or works, without adding a new capability and without correcting a defect. When one prompt introduces a new functionality, the sub-requests that together describe it, including ones that add detail to how it should look or work, are each Introduced, even when several of them fall in the same group; a sub-request that adds a further capability on top of a functionality already described or built is Expanded. Hedged wording can accompany either and does not by itself decide the code.
- Repaired: the sub-request says that something that was requested, or that the agent built, does not work, is broken, or behaves differently from what was requested, and wants it corrected. It is Repaired when the prompt itself reports the problem, whatever the form of the sentence; but when it asks to fix the same problem as a sub-request listed before it in the same prompt, adding nothing of its own, it is Restated. A complaint that the software does not do something nobody requested or built asks for something new; code it by what it asks for.
- Restated: the sub-request repeats a request already made, in an earlier prompt or by a sub-request listed before it in the same prompt, in the same or in different words. When two sub-requests of the same prompt ask for the same thing (the same feature, or the fix of the same problem), the first gets its own code and the later one is Restated. A sub-request that asks again for an earlier prompt's request is Repaired when the prompt says that it was not done or does not work, and Restated when the prompt does not say so. If it adds a requirement the earlier request did not ask for, changing how that functionality should look or work, it is Refined; wording that only spells out the same request more fully is still Restated. A sub-request that only asks for the software as the sub-requests listed before it already describe it, adding nothing of its own, is also Restated; one that asks for a platform or form of the software different from anything earlier asked for is coded as that new request, and one that only asks how to do that is Information. An umbrella request that comes before the sub-requests it summarises repeats nothing and is coded by what it asks for.
- Narrowed vs Abandoned. Narrowed keeps a feature but restricts where, for whom or to what it applies; moving or rearranging it without reducing it is Refined. Abandoned removes or drops something the software has or was asked to have, a whole functionality or a separate feature within one (including one the agent built without being asked), even when the rest of its group stays; so Abandoned can be the code of a group's first sub-request.
- Information vs Verification. Information asks the agent to explain, instruct, locate or recommend, without asking it to implement anything: what something is or is for, what it does or why, how to do something with the software, where something is, what to do next. Verification asks the agent to check or confirm, and report back, that a particular functionality, change or result is in place or works. A request that the software be made to behave a certain way asks for implementation and is coded by the change it asks for, unless it only asks the agent to check and report. A question that only asks whether something works or is in place, without the prompt reporting a problem, is Verification or Information.
\end{lstlisting}

\paragraph{User message.}\mbox{}

\begin{lstlisting}[style=prompt]
Functionality groups of this participant's software:

Each group is listed with all of its sub-requests over the whole session, marked as earlier than, in, or later than the prompt being coded.

G1 <group name>
    <sub-request ID> (earlier) <sub-request>
    <sub-request ID> (THIS PROMPT) <sub-request>
    <sub-request ID> (later) <sub-request>
G2 <group name>
    …

Earlier prompts in this session, oldest first, each with its sub-requests:

[<line ID>] <h:mm:ss> PROMPT 1: <prompt>
    <sub-request ID> [G<n> <group name>] <sub-request>
[<line ID>] <h:mm:ss> PROMPT 2: <prompt>
    …

Transcript from the previous prompt to the next prompt (think-aloud and interview; the AI agent's replies are not in it). The prompt being coded is marked >>>.

[<line ID>] <h:mm:ss> PARTICIPANT: <text>
>>> [<line ID>] <h:mm:ss> PROMPT: <prompt>
[<line ID>] <h:mm:ss> RESEARCHER: <text>
…

Prompt being coded (PROMPT <k>, [<line ID>]):
<prompt>

Its sub-requests, in order:
<sub-request ID> [G<n> <group name>] <sub-request>
…

Give each of these sub-requests one evolving code.
\end{lstlisting}

For the first prompt, the list of earlier prompts reads
\texttt{(none: this is the first prompt)}, and the transcript starts at the
prompt itself.

\subsection{On-Screen Action and Verification Prompts}
\label{sec:appendix-prompt-actions}

The model labels each 2-second frame of the building phase with one action.
For a frame labeled \emph{testing / reading}, it also names what the participant tests or
reads: the agent's output, one of the participant's functionality groups, or a
functionality outside the groups. A functionality counts as verified when at
least one frame names it. The model and settings are those of the planning and
evolution prompts: \texttt{qwen3.6-35b-a3b-q6} at temperature 0 with thinking
off, returning JSON constrained by a strict schema. Each frame is labeled in a
single run. \cref{tab:llm-action-labels} gives the name used in the paper for
each label.

\begin{table}[htbp]
\centering
\footnotesize
\caption{On-screen action labels and their names in the paper.}
\label{tab:llm-action-labels}
\begin{tabular}{@{}ll@{}}
\toprule
Label & Name in the paper \\
\midrule
\texttt{testing\_reading} & testing / reading \\
\texttt{waiting} & waiting \\
\texttt{prompting} & prompting \\
\texttt{configuring} & configuring \\
\texttt{editing\_code} & editing code \\
\texttt{other\_window} & other window \\
\texttt{no\_screen} (set without a model request) & no screen \\
\bottomrule
\end{tabular}
\end{table}

\paragraph{System message.}
The participant's functionality groups replace the lines in angle brackets.

\begin{lstlisting}[style=prompt]
You are coding a screen recording of somebody building an application for
their own use with an AI coding agent, one label per screenshot.

Screenshots are 2 seconds apart and given in order, each stamped with its time.
Assign exactly one label to every screenshot.

Each stamp also says how much of the screen changed in the second before the
screenshot and in the second after it, measured on the pixels: over the whole
screen, then in its left, middle and right thirds. Under 0.5% is a still screen
apart from the pointer or a timer; more is a scroll, a new page or dialog, typed
or streaming text, or something redrawn. Use it to see what moves, then look at
the screenshots to tell who moved it: the person (scrolling, clicking, typing,
navigating) or the agent (streaming its reply, listing files, reloading the
preview after an edit).

Two surfaces are on screen through most of the session and must not be confused:
the TOOL (the agent's chat or terminal session, its replies and changelog, its
code editor, its settings) and the APPLICATION being built (in a preview pane, a
browser tab or its own terminal). Acting in the application is working on the
application; acting in the chat is working with the agent.

For every screenshot, first say where things are, in thirds of the screen's
width (left, middle, right, left and middle, middle and right, whole screen, or
not visible):
  tool_area    the chat or terminal panel of the agent session the application
               is built in
  code_editor  a code editor: a source file open for editing, in the tool's code
               view, an IDE or a text editor. Code shown in the agent's reply, a
               diff or change the agent presents, and a terminal are not a code
               editor.
and what the mouse pointer is on:
  pointer      application, tool, code editor, elsewhere, or not visible
Read the stamps' left, middle and right figures against these areas to see which
part of the screen changed.

Then read agent_status from what the screenshot shows:
  working            a live sign that the agent is running: a timer or step
                     counter still advancing, a spinner or an "...ing" status,
                     output still streaming, files still being edited
  finished           it has stopped: its reply is complete, with a final time or
                     a done mark, or it ended in an error
  asking_the_person  it waits for the person to approve or choose before it goes
                     on
  not_visible        no agent status on screen

Then give the screenshot its code. Each code is defined below; an activity lasts
from its start to its end, and a screenshot takes the activity in progress at its
time.

  testing_reading  The person tests the application or reads what is on screen.
                   Testing is operating the application -- clicking its controls,
                   entering data, scrolling or navigating its pages, opening its
                   dialogs, running its commands -- at any time, also while the
                   agent works; the agent changing or reloading the application
                   is not the person testing it. Reading is looking at the
                   application or at what the agent wrote: its reply, changelog,
                   the code it wrote, a question it asks. Once the agent has
                   stopped, whatever the person looks at is reading, even a still
                   screen. While the agent works, only real reading counts: the
                   person stays on something and scrolls it or follows it with the
                   pointer, through the pauses between scrolls, judged over about
                   5 seconds around the screenshot.
  editing_code     The person edits code by hand in a code editor: a code editor
                   is on screen (code_editor is not "not visible"), the caret is
                   in it, and characters the person types appear or disappear in
                   its code. Typing into the application's own fields is
                   testing_reading, and typing a file name in a file list is not
                   editing code either. Scrolling, selecting or reading code is
                   not editing, and code the agent writes, changes or shows for
                   approval never counts, even inside the editor.
  prompting        The person writes a message to the agent: the caret is in the
                   agent's chat input, from the first character typed, through
                   pauses mid-draft, until the message is sent, also while the
                   application or the agent's reply is on screen.
  configuring      The person works in the agent session's own settings: its
                   model, API keys, sign-in, sharing, billing or deployment, for
                   as long as such a panel or dialog is open. The application's
                   own settings are testing_reading.
  other_window     Neither the agent session nor the application is in front:
                   another window, tab or website, or a part of the tool other
                   than the session in which the application is built, whatever
                   the person does there.
  waiting          The agent is working and the person is not testing, reading,
                   editing, writing a message or configuring: they watch its
                   output stream or its timer count, move the pointer here and
                   there without settling on anything, or do nothing. It lasts
                   from when the agent starts, usually when a message is sent,
                   until it stops or the person starts another activity.

When the label is testing_reading, `target` names what the person is testing or
reading, as the screen itself labels it -- the application's screen, tab,
control or feature, or the agent's message. Use the on-screen wording, not a
paraphrase. `functionality` then says which it is:
  chat              the agent's output in the tool, which the person is reading:
                    its reply, changelog, plan, the code it wrote, its approval
                    request
  G1, G2, ...       the application, when the target is the feature a group is
                    named for, or a part of it: its screen, form, chart, tab or
                    control. The group's name and requests say what the feature
                    is. A page that merely contains the feature counts only while
                    the person is using that feature.
  not in the list   the application, when the target is a feature no group is
                    named for: a screen or control no request asks for, or a main
                    page looked over as a whole
When the person operates the application it is never chat. When they only look,
it is what they are looking at, judged by where the pointer is and what has just
changed: the agent's output (chat) or the application's feature. The other codes
have no target or functionality.

The person's requests to the agent, grouped by feature:
  G1 <group name>: <request>; <request>; …
  G2 <group name>: <request>; …
  …

Judge each screenshot on what it shows, using the ones around it only to resolve
which of two labels a moment belongs to -- whether the screen changed, whether a
timer advanced. `evidence` is a short clause naming what in the image decided it:
a visible timer, the caret's position, a control being operated, the screen being
unchanged from the previous frame. Do not infer intent or explain why the person
is doing it.
\end{lstlisting}

\paragraph{User message.}
The message is a list of text and image parts: one opening line, then a text
stamp and an image for each frame.

\begin{lstlisting}[style=prompt]
<number of frames> screenshots, 2s apart, in order. Label every one.

t=<seconds>s (<h:mm:ss>), changed in the second before: <x>% (left <x>%, middle <x>%, right <x>%), changed in the second after: <x>% (left <x>%, middle <x>%, right <x>%):
<image: the frame, JPEG, 1,024 pixels wide>
…
\end{lstlisting}

In each stamp, \texttt{t} is the frame's time in seconds from the start of the
recording. The percentages have one decimal place, for example
\texttt{changed in the second before: 1.2\% (left 0.0\%, middle 3.1\%, right 0.5\%)}.
The before or after part is left out when there is no frame one second before
or after.

\section{Additional Discussion}
\label{sec:appendix-discussion}

\subsection{Authority Drift Is Not Automation Bias}

A substantial literature has studied automation bias---the tendency to
over-rely on automated suggestions even when they are
incorrect~\cite{parasuraman_humans_1997}---and the conditions under which people
appropriately rely on, override, or defer to AI
recommendations~\cite{lai_towards_2021, bansal_does_2021}. In creative contexts,
AI suggestions narrow the design space by anchoring creators on what the system
proposes~\cite{wadinambiarachchi_effects_2024, anderson_homogenization_2024}.

The authority drift we observed is structurally different. It is not a human
rubber-stamping a recommendation they can see, but an AI making decisions the
human does not encounter until later, if at all. Every detail a prompt leaves
open is one the agent must close to produce output. Seven of the eight
participants encountered functionality the AI had added without being asked,
five encountered requirements that were misunderstood or insufficiently
clarified, and three had difficulty determining what the AI was doing or had
changed.

This mechanism---underspecification as the transfer of authority---differs from
both automation bias, where the human sees the suggestion and fails to override
it, and the anchoring documented in creative AI tools, where the human sees the
options and fails to explore beyond them. It is closer to a contractor
interpreting an ambiguous blueprint, except that the contractor's interpretation
is buried in generated code the client cannot read.

\subsection{Mapping Observations to Learning Barriers}

\cref{tab:barrier-mapping} details the evidence behind our comparison with
the learning barriers of Ko, Myers, and Aung~\cite{ko2004six}. Counts are numbers of
participants unless noted.

\begin{table}[htbp]
\centering
\footnotesize
\setlength{\tabcolsep}{3pt}
\caption{Observed opportunities and difficulties mapped to the learning
barriers of Ko et al.~\cite{ko2004six}. A dash indicates no corresponding observation.}
\label{tab:barrier-mapping}
\begin{tabular}{@{}l>{\raggedright\arraybackslash}p{4.2cm}>{\raggedright\arraybackslash}p{5.4cm}@{}}
\toprule
Barrier & Opportunity observed & Difficulty observed \\
\midrule
Design & AI expanded design options (6) & 89 of 131 functionalities emerged during building; unwanted AI additions (7) \\
Selection & Useful technical assistance (7) despite limited app-development knowledge (4) & --- \\
Coordination & Agent chose and wired components; 0.2\% of building intervals spent editing code by hand & Service-integration difficulties (2); data-storage or access problems (3) \\
Use & --- & Repeated or reformulated requests (6); misunderstood requirements (5) \\
Understanding & --- & Mismatch between expected and delivered behavior (8) \\
Information & --- & Difficulty determining what the AI did (3); incorrect AI information (2); 38 of 131 functionalities without observed verification \\
\addlinespace
\emph{Stewardship} (beyond taxonomy) & --- & Uncertainty about future maintenance (3); limited ability to manage the generated software (2) \\
\bottomrule
\end{tabular}
\end{table}

\emph{Repaired} was the second most common request type and occurred for 37 of the 131
functionalities. Repairs proceeded through the same channel that had
produced the problem: another prompt to the agent.

\subsection{The Role of Expertise}

Programming experience, in our observation sample, did not straightforwardly
improve the creation process: participants with more programming background did
not produce noticeably better specifications. Two sessions instead suggest the
importance of verification habits from professional life. P4, a product designer
and chief of staff, realized the need for a backend and requested it in the first and second prompts, but discovered the absence of a backend only during the scenario
walkthrough, when closing and refreshing the tab reset everything. P7, a software engineer, was the only participant who
wrote testing criteria into prompts and asked the agent to write tests. What
distinguishes these participants is not that they knew what a backend or a test
suite was, but that they brought habits of verification from other domains of
their professional lives. Engineering support for personal software therefore
cannot assume a particular technical background, but it may be able to scaffold
the verification habits that experienced practitioners bring. Given our sample,
we treat this as a hypothesis for future work.

\subsection{Participant Reflections on Ownership}

The participant who rated ownership lowest (1 of 5) rated personal fit 4.33 and
stopped using the software, explaining that they wanted \pquote{an app that I have sense of ownership over}{P2}. Another participant, asked whether the experience had
changed how they would build software for themselves, said they would \pquote{rather
build the software myself with some aid from AI/Google rather than have the AI
build the entire software for me, especially if it is meant to be a personal
project}{P3}.

\end{document}